\documentstyle [12pt,epsfig]{article} 
\makeatletter
\@addtoreset{equation}{section}
\makeatother

\renewcommand{\theequation}{\thesection.\arabic{equation}}
\newcommand{\ba}{\begin{eqnarray}}
\newcommand{\ea}{\end{eqnarray}}

\newcommand{\A}{{\cal A}}
\newcommand{\Ei}{{\rm E}}
\newcommand{\Bi}{{\rm B}}
\newcommand{\Ai}{{\rm A}}
\newcommand{\Ji}{{\rm J}}

\newcommand{\Jnr}{{\it J}}

\begin{document}
\newcommand{\BS}{\bigskip}
\newcommand{\SECTION}[1]{\BS{\large\section{\bf #1}}}
\newcommand{\SUBSECTION}[1]{\BS{\large\subsection{\bf #1}}}
\newcommand{\SUBSUBSECTION}[1]{\BS{\large\subsubsection{\bf #1}}}

\begin{titlepage}
\begin{center}
\vspace*{2cm}
{\large \bf Retarded electric and magnetic fields of a moving
   charge: Feynman's derivation of Li\'{e}nard-Wiechert potentials
   revisited}
\vspace*{1.5cm}
\end{center}
\begin{center}
{\bf J.H.Field }
\end{center}
\begin{center}
{ 
D\'{e}partement de Physique Nucl\'{e}aire et Corpusculaire
 Universit\'{e} de Gen\`{e}ve . 24, quai Ernest-Ansermet
 CH-1211 Gen\`{e}ve 4.}
\end{center}
\begin{center}
{e-mail; john.field@cern.ch}
\end{center}
\vspace*{2cm}
\begin{abstract}
  Retarded electromagnetic potentials are derived from Maxwell's equations
  and the Lorenz condition. The difference found between these potentials
   and the conventional Li\'{e}nard-Wiechert ones is explained
   by neglect, for the latter, of the motion-dependence of
   the effective charge density. The corresponding retarded fields of a point-like charge
    in arbitary motion are compared
     with those given by the formulae of Heaviside, Feynman, Jefimenko and other authors. 
    The fields of an accelerated
    charge given by the Feynman are the same as those derived from the Li\'{e}nard-Wiechert potentials
   but not those given by the Jefimenko formulae. A mathematical error concerning
      partial space and time derivatives in the derivation of the Jefimenko 
      equations is pointed out.
\end{abstract}
\vspace*{1cm}{\it Keywords};  Special Relativity, Classical Electrodynamics.
\newline
\vspace*{1cm}
\end{titlepage} 
   
\SECTION{\bf{Introduction}}
  The present paper is the fifth in a series written recently by the present author on relativistic classical
  electrodynamics (RCED). In the first of the papers ~\cite{JHFRCED}, all of the formulae of classical 
  electromagnetism (CEM), up to relativistic corrections of O($\beta^2$), relating to intercharge forces,
  were derived from Hamilton's Principle, assuming only Coulomb's inverse-square force law of electrostatics
   and relativistic invariance. In the same paper it was shown that the intercharge
   force, mediated by the exchange of space-like virtual photons, is
  predicted by quantum electrodynamics (QED) to be instantaneous in the centre-of-mass frame of the interacting charges.
   Recently, convincing experimental evidence has been obtained~\cite{Kohletal} for the non-retarded nature
   of `bound' magnetic fields with $r^{-2}$ dependence, (associated in QED with virtual photon exchange)
   in a modern version, probing small $r$ values, of the Hertz experiment~\cite{Hertz} in which the electromagnetic
   waves associated with the propagation of real photons (fields with $r^{-1}$ dependence) were
   originally discovered. 
     \par In two subsequent papers~\cite{JHFRSKO,JHFIND} the predictions of the RCED formulae for intercharge
   forces derived in Ref.~\cite{JHFRCED} are compared with the predictions of the CEM (Heaviside)
   formulae~\cite{Heaviside} for the force fields of a uniformly moving charge. Unlike the RCED formulae,
   the CEM ones correspond to
   a retarded interaction. If the latter are written in `present time' form~\cite{PPPT} they are found to
   differ from RCED formulae by terms of O($\beta^2$). In the first paper~\cite{JHFRSKO}, it is shown that
   consistent results for small-angle Rutherford scattering in different inertial frames are obtained only
   for RCED formulae and that stable, circular, Keplerian orbits of a system consisting of two equal and
   opposite charges are impossible for the case of the retarded CEM fields. The related `Torque Paradox' of
   Jackson~\cite{JackTP} is also resolved by use of the instantaneous RCED fields.
     The second paper~\cite{JHFIND} considers electromagnetic induction in different reference frames.
    Again, consistent results are obtained only in the case of RCED fields. It is demonstrated that for
    a particular spatial configuration of a simple two-charge `magnet' the Heaviside formula for the
    electric field predicts a vanishing induction effect in the case that the `magnet' is in motion and the
    test coil is at rest.
    \par In Ref.~\cite{JHFFT}, the space-time transformation properties of the RCED and CEM force fields
    were studied in detail and compared with those that provide the classical description of the creation, 
    propagation, and destruction of real photons. It was shown that in the relativistic theory longitudinal
    (with respect the direction of motion of the source charge) electric fields contain covariance-breaking
    terms of O($\beta^2$). The electric Gauss Law and Electrodynamic (Amp\`{e}re Law) Maxwell Equations
    are also modified by the addition of covariance-breaking terms of  O($\beta^4$) and O($\beta^5$)
    respectively. The retarded fields are re-derived from the Maxwell Equations and the
    Lorenz condition and an error in the derivation
    of the retarded Li\'{e}nard-Wiechert (LW)~\cite{LW} potentials was pointed out. The argument leading to
    this conclusion ---which implies that retarded fields given by the
    Heaviside formulae are erroneous for this trivial mathematical
    reason, as well as being inconsistent with QED--- is recalled in Sections 2 and 3 below. 
     \par The aim of the present paper is to present a more detailed discussion of retarded electromagnetic
       fields with a view to pointing out some of the mathematically erroneous statements on this
        subject that have appeared in classical research literature, text books and modern
        pedagogical literature. The correct relativistic formulae for the retarded fields of
        an accelerated charge have previously been derived in Ref.~\cite{JHFFT}. These fields
        actually describe only the production and propagation of real photons whereas in text books
        and the pedagogical literature it is universally assumed that these fields describe both
       intercharge forces and radiative effects. Since the present paper is concerned only
       with the postulates and mathematical arguments used in different derivations of the
     retarded fields, the physical interpretation of the fields (in particular their relation
       to the quantum mechanical description of radiation), as discussed in Ref.~\cite{JHFFT},
        is not considered.
  \par The structure of the paper is as follows. In the following section the retarded 4-vector
      potential is derived from inhomgeneous d'Alembert equations and the Lorenz condition.
       The reason for the difference between the potential so-obtained and the pre-relativistic
        LW potentials is explained. In Section 3 Feynman's derivation of the LW potentials
       is recalled, where the `multiple counting' committed also in the original derivations~\cite{LW}
        is made particularly transparent. In Section 4 some erroneous `relativistic' derivations
        of the LW potentials and the Heaviside formulae that are commonly presented in text books
          on classical electromagnetism are discussed. In Section 5 the retarded fields of a
         uniformly moving charge are considered and the `present time' formulae for the retarded
        RCED fields are derived for comparison with the Heaviside formulae of CEM. In  Section 6
        a comparison is made between different formulae for the retarded fields of an accelerated charge 
        that have appeared in text books and the pedagogical literature including the well-known
        ones of Feynman and Jefimenko. Section 7 contains a brief
  summary.

\SECTION{\bf{Derivation of retarded electromagnetic potentials from inhomogeneous
 d'Alembert equations}}
 
 As described in Ref.~\cite{Jack1}, retarded electromagnetic potentials may be derived from 
 the Maxwell equations:
    \begin{eqnarray}
      \vec{\nabla} \cdot \vec{\Ei} & = & 4 \pi \Ji_0, \\
       \vec{\nabla} \times \vec{\Bi} & - &\frac{1}{c}
      \frac{\partial \vec{\Ei}}{\partial t} =  4 \pi \vec{\Ji} 
    \end{eqnarray}
   and the Lorenz condition
   \begin{equation}
  \vec{\nabla} \cdot \vec{\Ai} + \frac{1}{c}\frac{\partial \Ai_0}{\partial t}  = 0 
    \end{equation} 
     where the current density $\Ji$ is a 4-vector:
   \begin{equation}
 \Ji(\vec{x}_J(t),t) = (\Ji_0;\vec{\Ji}) \equiv (\gamma_u \rho^*; \gamma_u \vec{\beta}_u \rho^*) = \frac{u  \rho^*}{c}.
    \end{equation} 
   The system of source charges is assumed to be at rest in the frame S$^*$, where the charge density
   is $\rho^*$, and to move with velocity $\vec{u} = c \vec{\beta}_u$ relative to the frame S in which the potential
   is defined. The 4-vector velocity of the charge system in this last frame is:
   \begin{equation}
     u \equiv (c\gamma_u ; c\gamma_u \vec{\beta}_u )
    \end{equation} 
   where
    \[ \beta_u \equiv \frac{u}{c},~~~\gamma_u \equiv \frac{1}{\sqrt{1-\beta_u^2}}. \] 
 The first step of the calculation is to use the Lorenz condition (2.3) to eliminate either $\vec{\Ji}$ or
     $\Ji_0$ from (2.1) and (2.2) to obtain the inhomogeneous d'Alembert equations:
        \begin{eqnarray}
      \nabla^2 \Ai_0 -\frac{1}{c^2}\frac{\partial^2 \Ai_0}{\partial t^2} & = & -4 \pi \Ji_0,  \\
       \nabla^2 \vec{\Ai} -\frac{1}{c^2}\frac{\partial^2  \vec{\Ai}}{\partial t^2} & = & -4 \pi \vec{\Ji}.
       \end{eqnarray}
         These equations are readily solved by introducing appropriate Green functions~\cite{Jack1}. The
       solutions give the retarded 4-vector potential:
    \begin{equation}
     \Ai_{\mu}^{ret}(\vec{x}_q,t)  = \int dt' \int d^3 x_J(t') \frac{\Ji_{\mu}(\vec{x}_J(t'), t')}
       {|\vec{x}_q -\vec{x}_J(t')|} \delta(t'+\frac{|\vec{x}_q -\vec{x}_J(t')|}{c}-t).
     \end{equation} 
       Here $\vec{x}_q$ is the position and $t$ the time at which the potential is defined and $\vec{x}_J(t')$ specifies
   the position of the volume element $d^3 x_J(t')$ at the earlier time $t'$. The $\delta$-function ensures that the volume element
    lies on the backward light cone of the field point specified by  $\vec{x}_q$, as required by causality,
     since the potentials give the classical description of the propagation, from the source to the field
      point, of real (on-shell) photons at speed $c$. This is a consequence of the wave-equation-like
       structure of the terms on the left sides of the  d'Alembert equations. The solutions of
      the corresponding homogeneous d'Alembert equations are progressive waves with phase
       velocity $c$. 
       \par In the special case of a single point-like source charge the current density in (2.8)
  is given by the expression:
  \begin{equation}
    \Ji^Q(\vec{x}_J(t'), t') = \frac{ Q u}{c} \delta (\vec{x}_J(t')-\vec{x}_Q(t')) 
     \end{equation}
     where $\vec{x}_Q(t')$ is the position of the charge at time $t'$.
    Inserting (2.9) in (2.8), and integrating over $\vec{x}_J$, gives
     \begin{equation}
     \Ai_{\mu}^{ret}(\vec{x}_q,t)  = \frac{Q u_{\mu}}{c} \int dt'  \frac{\delta(t'-t'_Q)}
       {r'}
     \end{equation}
      where
  \begin{equation}
     r' \equiv |\vec{x}_q-\vec{x}_Q(t')|,~~~ t'_Q \equiv t - \frac{|\vec{x}_q -\vec{x}_Q(t'_Q)|}{c}
    = \left. t - \frac{r'}{c} \right|_{t' = t'_Q}. 
  \end{equation}
 The retarded 4-vector potential is therefore:
   \begin{equation}
  (\Ai_0^{ret};\vec{\Ai}^{ret}) = \left( \left. \frac{Q \gamma_u}{r'} \right|_{t' = t'_Q};
     \left. \frac{Q \gamma_u \vec{\beta}_u}{r'}\right|_{t' = t'_Q} \right).
   \end{equation}
   
   \par A similiar result to (2.12) is obtained in the case of an extended distribution of charge
    in the case that its dimensions are much less than the separation between the average position
     of the source charge distribution, $\langle \vec{x}_J\rangle$, and the field point. In this case
     $\vec{x}_J$ may be replaced in the $\delta$-function and denominator of
     (2.8) by $\langle \vec{x}_J\rangle$, so that the
      factor $\langle r' \rangle \equiv |\vec{x}_q -\langle \vec{x}_J \rangle|$ in the denominator
      may be taken outside the 
       $\vec{x}_J$ integral giving
    \begin{eqnarray}
    \Ai_{\mu}^{ret}(\vec{x}_q,t) & = & \int\frac{dt'}{\langle r' \rangle} \int d^3 x_J( t'_J)
       \Ji_{\mu}(\vec{x}_J(t'),t')
    \delta(t'+\frac{|\vec{x}_q - \langle \vec{x}_J\rangle|}{c}-t) \nonumber \\  
       & = & \int\frac{dt'}{\langle r' \rangle}\frac{u_{\mu}}{c} \int d^3 x_J(t') \rho^*(\vec{x}_J(t'), t')
      \delta(t'+\frac{|\vec{x}_q -\langle \vec{x}_J\rangle|}{c}-t) \nonumber \\  
       & = &  \int\frac{dt'}{\langle r' \rangle} \frac{u_{\mu} Q }{c} \delta(t'-\langle t'_J \rangle)
      \end{eqnarray}
      where $Q$ is the total charge of the distribution:
       \begin{equation}
          Q =   \int  \rho^*(\vec{x}_J(t'), t') d^3 x_J( t') 
       \end{equation} 
      and 
    \begin{equation}
  \langle t'_J  \rangle \equiv t - \frac{|\vec{x}_q -\langle \vec{x}_J \rangle|}{c} 
   =  t - \frac{\langle r' \rangle}{c}  
  \end{equation} 
           giving the 4-vector potential:
     \begin{equation}
  (\Ai_0^{ret};\vec{\Ai}^{ret}) = \left( \left. \frac{Q \gamma_u}{\langle r' \rangle}
      \right|_{t' = \langle t'_J  \rangle};
     \left. \frac{Q \gamma_u \vec{\beta}_u}{ \langle r' \rangle}\right|_{t' = \langle t'_J \rangle} \right).
   \end{equation}
   \par It is now of interest, in view of understanding the origin of the LW potentials, to
    recalculate the retarded potentials after inverting the the order of the $t'$ and
     $\vec{x}_J(t')$ integrations in (2.8), so that:
    \begin{equation}
     \Ai^{ret}_{\mu}(\vec{x}_q,t)  = \int d^3 x_J(t') \int dt' \frac{\Ji_{\mu}(\vec{x}_J(t'), t')}
       {|\vec{x}_q -\vec{x}_J(t')|} \delta(t'+\frac{|\vec{x}_q -\vec{x}_J(t')|}{c}-t).
     \end{equation} 
      Unlike in (2.10), where the insertion of the current density of a point-like charge, (2.9) simply specifies
   the value of $t'$ in the $\delta$-function to be $t'_Q$, as given by Eq.(2.11), on integrating over $\vec{x}_J$,
    the argument of the $\delta$-function in (2.17) has a more complicated dependence on $t'$:
 \begin{equation}
     \delta [f(t')] = \frac{\delta(t'-t'_J)}{~~~~\left|\frac{\partial f(t')}{\partial t'}\right|_{t '= t'_J}}
     \end{equation} 
   where $t'_J$ is the solution of the equation $f(t')=0$ and
 \begin{equation}
    f(t') \equiv t' +\frac{|\vec{x}_q -\vec{x}_J(t')|}{c}-t. 
   \end{equation}
 It follows from (2.19), and the definition of $t'_J$, that
 \begin{equation}
     t'_J =  t - \frac{|\vec{x}_q -\vec{x}_J(t'_J)|}{c}.   
  \end{equation}
  Differentiating (2.19) gives:
\begin{equation}
   \frac{\partial f(t')}{\partial t'} = 1-\hat{r}'_J \cdot \vec{\beta}_u
  \end{equation}
  where:
     \begin{equation}
     \hat{r}'_J = \frac{\vec{x}_q -\vec{x}_J(t_J')}{|\vec{x}_q -\vec{x}_J(t_J')|},~~~
      \vec{\beta}_u = \frac{1}{c}\frac{d \vec{x}_J(t')}{d t'}
     \end{equation}
  so that (2.17) may be written as
     \begin{equation}
     \Ai^{ret}_{\mu}(\vec{x}_q,t)  = \int d^3 x_J(t') \int dt' \frac{\Ji_{\mu}(\vec{x}_J(t'), t')}
       {|\vec{x}_q -\vec{x}_J(t')|(1-\hat{r}'_J \cdot \vec{\beta}_u)} \delta(t'- t_J').   
     \end{equation} 
 In performing the integral over $t'$, proper account must now be taken of the appropriate
 current density $J_{\mu}$ to be inserted in (2.23). The limits of the $t'$ integral 
  are determined by the times at which the backward light cone of the field point
   coincides with the boundaries of the moving charge distribution. This is illustrated
   in Fig.1 for a uniform block of charge DEFG, of trapeziodal shape, moving in the plane
    of the figure towards a distant field point, in this plane, far to the right. 
    The segments AA', BB' and CC' lie on the light front, LF, that coincides with the
    backward light cone of the field point. It is assumed that the latter is sufficiently
    far that LF may be approximated by a plane, with normal in the plane of the figure.
    The block of charge is moving with speed $u$ in the plane of the figure at angle
   $\theta$ to the direction of motion of LF. The light front starts to overlap the 
    block of charge in the position AA' and ceases to do so in the position CC'.
    The limits of the $t'$ integral in (2.23) for this case then correspond to 
     the times when the front coincides with AA' (lower limit) and with
     CC' (upper limit). Inspection of Fig.1 shows that, during the time interval between
     these limits, the average value of the charge density, $\bar{\rho}$, is less than
     that when the distribution it at rest, $\rho^*$, by the ratio:  
      \begin{equation}
   \frac{\ell}{L} = \frac{{ \rm length~of~charge~distribution}}{{\rm length~of~light~cone~overlap~region}}.    
    \end{equation}
    If $\Delta t'$ is the time during which there is overlap between LF and the block of charge,
    the geometry of Fig.1 gives:
      \begin{equation}
     L = u \Delta t' + \ell = \frac{c \Delta t'}{\cos \theta}
  \end{equation}
     so that
    \begin{equation}
   \frac{\ell}{L} = 1-\frac{u}{c} \cos \theta = 1-\hat{r}'_J \cdot \vec{\beta}_u.
  \end{equation}
    It can be seen from Fig.1 that the same average charge density is obtained
    if the uniform block of charge is replaced by a point-like charge, $Q$, equal
    to the integrated charge of the block and placed at its centre, or if the 
    moving uniform charge distribution is replaced by the fixed one MNOP with
    density $\bar{\rho}$. For a single point-like charge the appropriate
     current density in (2.23) is then given by (2.9). (2.24) and (2.26) as:
 \begin{equation}
    \Ji^Q(\vec{x}_J(t'), t') = (1-\hat{r}'_J \cdot \vec{\beta}_u)
  \frac{ Q u}{c} \delta (\vec{x}_J(t')-\vec{x}_Q(t')). 
     \end{equation}
   Inserting (2.27) in (2.23) and performing the integrals over $t'$ and $\vec{x}_J$ recovers the
   result of Eq.(2.12). The increased overlap time of the light front resulting from
   the motion of the block of charge is exactly compensated by the reduction of
  the average charge density resulting from the same motion. The incorrect LW potentials
  are given by taking into account the time-everlap correction factor but neglecting the
  corresponding change in the charge density. This gives, instead of (2.12), the potentials
    \begin{equation}
  (\Ai_0^{ret};\vec{\Ai}^{ret}) \equiv (\gamma_u \Ai_0({\rm LW})^{ret};\gamma_u\vec{\Ai}({\rm LW})^{ret})
   =  \left( \left. \frac{Q \gamma_u}{r'(1-\hat{r}'_J \cdot \vec{\beta}_u)} \right|_{t' = t'_Q};
     \left. \frac{Q \gamma_u \vec{\beta}_u}{r'(1-\hat{r}'_J \cdot \vec{\beta}_u)}\right|_{t' = t'_Q} \right)
   \end{equation}
    where $\Ai_0({\rm LW})^{ret}$ and $\vec{\Ai}({\rm LW })^{ret}$ are the  Li\'{e}nard and Wiechert potentials.
  This mistake in the original Li\'{e}nard and Wiechert~\cite{LW}
    calculations has been repeated in all text-book treatments of the subject of
    retarded potentials. Some examples may be found in Refs.~\cite{PPLW,RosserLW,JackLW,SchwartzLW,
    GriffithsLW}.
   \par Inspection of (2.23) shows that neglect of the charge density correction factor of (2.27) in
      evaluating the potentials implies that they are under-estimated when the source is receding
     ($\hat{r}'_J \cdot \vec{\beta}_u < 0$) and over-estimated when it is approaching 
      ($\hat{r}'_J \cdot \vec{\beta}_u > 0$). On the other hand, the $1/r'$ dependence of the potential
      implies that the potentials are greater (smaller) when $\hat{r}'_J \cdot \vec{\beta}_u < 0$
       ($\hat{r}'_J \cdot \vec{\beta}_u > 0$). It is shown in Section 5 below that for the `present time'
     LW fields (Eqs.(4.14) and (4.15) below) the neglect of the charge density correction factor
      results in exact compensation of the $1/r'^2$ dependence so that the magnitudes of the fields are
     independent of the sign of $\hat{r}'_J \cdot \vec{\beta}_u$, as is the case for an instantaneous
     intercharge interaction.
 
   \par Note that the potentials on the right side of (2.28), derived by neglecting
 the density correction factor in (2.27), differ from the retarded LW potentials
    by an overall factor of $\gamma_u$. This factor will be
    commented on at the end of Section 4 below where alternative `relativistic'  derivations of the LW potentials
    are discussed.

\SECTION{\bf{Feynman's derivation of the Li\'{e}nard-Wiechert potentials}}
  The erroneous nature of the retarded potentials found when the charge density
  correction factor of Eqs.(2.24) and (2.26) is neglected is made particularly clear by
   a careful examination of Feynman's derivation~\cite{FeynLW} of the LW potentials
  for the case of parallel motion of the source distribution and the light front, LF,
  corresponding to the backward light cone of the field point.

 \begin{figure}[htbp]
\begin{center}
\hspace*{-0.5cm}\mbox{
\epsfysize9.0cm\epsffile{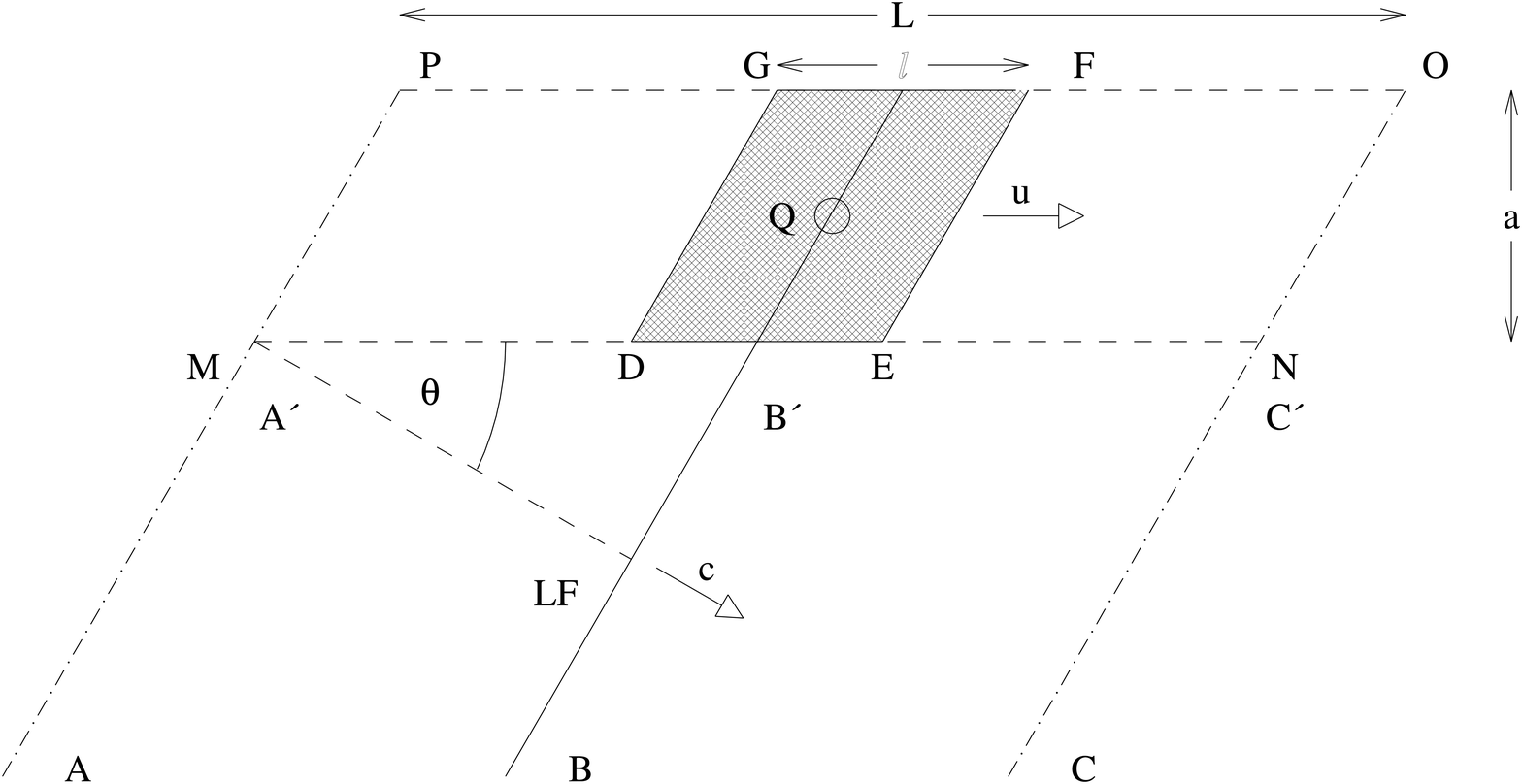}}   
\caption{{\sl The plane light front (LF) BB' crosses the block of charge
   DEFG with uniform charge density $\rho$ while moving from the position
 AA' to CC'.The light front and the block move in the plane of
  the figure with speeds $c$ and $u$ respectively in the directions 
   indicated. The average charge density sampled by the light 
   front during its passage over the block is $\bar{\rho} = \rho^* \ell/L$ The retarded
  potential generated by the charge of the block at a distant field
  point to the right of the figure is the same as that that would be
   generated by a block of charge in the form MNOP with the same
  depth as DEFG,with uniform charge density  $\bar{\rho}$, at rest, or
   by a moving point-like charge $Q =\rho^* V$,  where $V$ is the volume
   of the block DEFG.}}
\label{fig-fig1}
\end{center}
\end{figure}

\begin{figure}[htbp]
\begin{center}
\hspace*{-0.5cm}\mbox{
\epsfysize15.0cm\epsffile{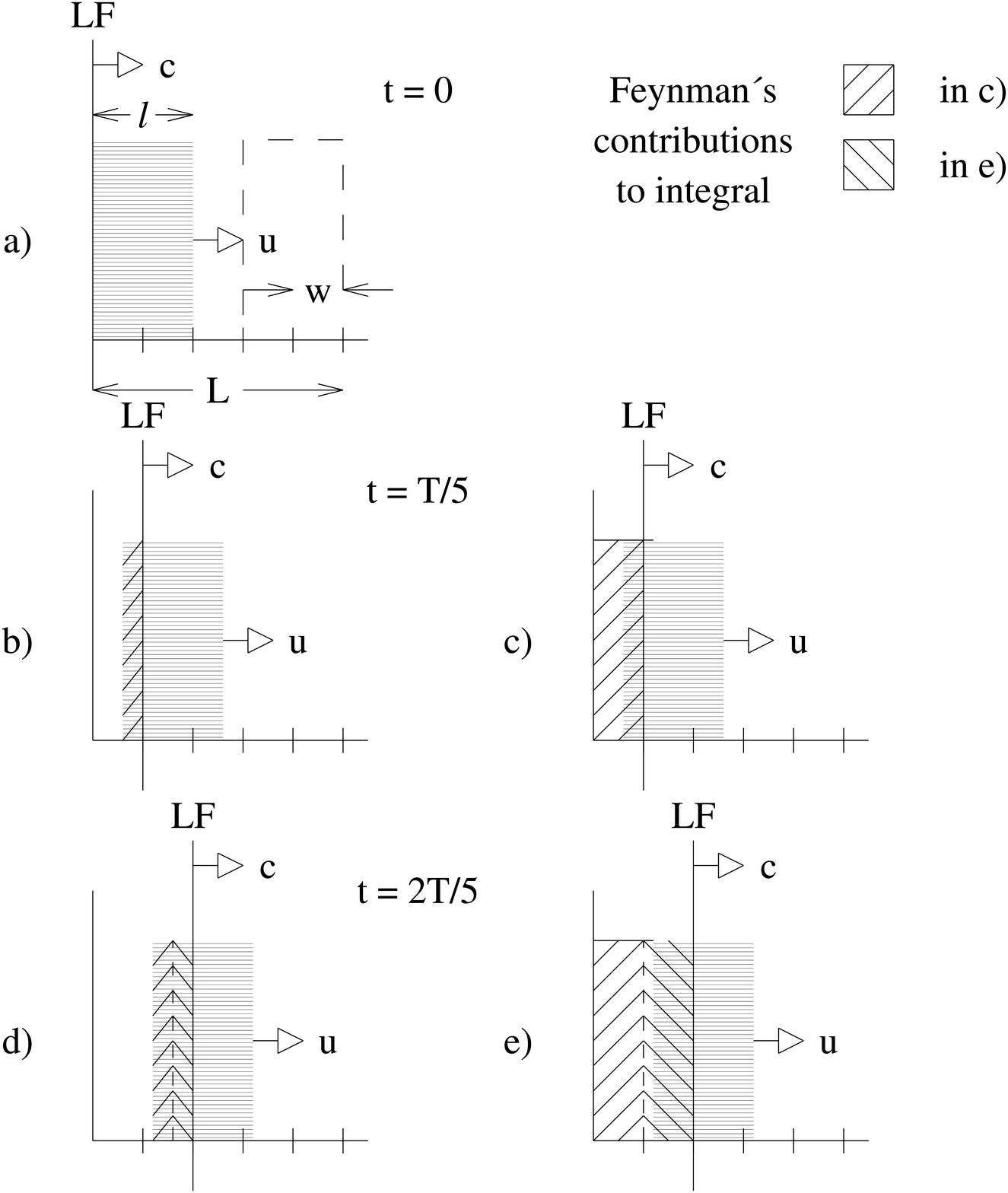}}   
\caption{{\sl Feynman's method of calculating retarded potentials~\cite{FeynLW}.
  A uniform rectangular block of charge of length  {\it l} moves to the right with speed $v$ towards
  a distant field point. The light front, LF, in causal connection with the field point, overlaps the block
  for a distance $L$ and a time $T$. In a) LF arrives at the front of the block. The position of the block
  when LF overtakes it is shown shaded. b) and d) show the positions of LF at times $t = T/5$ and $t = 2T/5$
  respectively. The regions of the block sampled by LF in the time intervals
  $0 < t < T/5$ and  $T/5 < t < 2T/5$ are shown by the SW-NE and NW-SE cross-hatched areas, respectively.
   The similar crossed-hatched areas in c) and e) show the charge volumes assigned to the potential 
    integral, during the same time intervals, in Feynman's calculation. See text for discussion.}}
\label{fig-fig2}
\end{center}
\end{figure}

     \par  Feynman's analysis of the problem of retarded potentials is shown in Fig.2. 
       A rectangular block of charge, of uniform density, moves towards the field point, which is sufficiently
     far to the right that the variation of $r'_J$ may, as in deriving Eq.(2.16) above,
     be neglected in evaluating the integral
     that gives the potential. The light front moves across the charge distribution, sampling it.
     Each element of charge which is crossed by LF gives a contribution to the potential. The depth of the 
    block of charge is $\ell$ and LF moves over the distance $L$ while crossing the charge 
    distribution. The front overlaps the charge distribution for a time interval $T$. The 
     overlap distance, $L$, is divided into bins of width $w$ and the contribution to the potential
    of each bin is considered separately. In Fig.2 the dimensions and velocity $u$ are chosen so that:
     \[ \ell = \frac{2 L}{5},~~~~~w = \frac{L}{5}. \]
     It then follows that $u = 3c/5$. In this figure, the positions of the charge distribution
     and the front LF at times 0, $T/5$, $2T/5$ respectively are shown. In Figs.2b,2d the front
    has crossed charge thicknesses of $0.4 w$,$0.8w$ respectively. The region crossed during the time
    $0 < t < T/5$ is shown by SW-NE\footnote{The points of the compass: South-West (SW), North-East (NE), 
     North-West (NW) and South-East (SE).}diagonal cross-hatching, that crossed in   $T/5< t < 2T/5$
      by NW-SE diagonal cross-hatching. Thus the average charge density in each bin is reduced, in comparison
    with the situation when the charges are at rest, by 60$\%$. Integrating first over the time,
      as in (2.17), for each bin, then gives:
  \begin{equation}
    \Ai_{\mu} = \frac{u_{\mu} S}{c r'_J} \sum_{bins}w \bar{\rho} = \frac{u_{\mu} S L \bar{\rho}}{c r'_J}
   \end{equation}
 where $S$ is the surface area of the charge distribution normal to its direction of motion and
   $\bar{\rho}$ is the average charge density. From the geometry of Fig.2a,  $\bar{\rho} = 2 \rho^*/5$ where 
   $\rho^*$ is the rest frame charge density. Since $L = 5 \ell /2$ (3.1) gives: 
 \begin{equation}
    \Ai_{\mu} = \frac{u_{\mu} S \ell \rho^*}{c r'_J} = \frac{u_{\mu} Q}{c r'_J}
   \end{equation}
    where $Q$ is the total charge in the block. Allowing for the propagation time delay of the light front
   with respect to the time of the field point (3.2) agrees with Eq.(2.16) but not with the LW potentials
   in (2.28).
    \par The contributions to the integral given by the first two bins, according to Feynman's original
    calculation~\cite{FeynLW} are shown by the SW-NE and NW-SE diagonal hatching in Figs.2c and 2e respectively.
    The movement of the charge distribution is neglected, and with it the change in the effective
     charge density. Feynman's result is given by replacing $\bar{\rho}$ in (3.1) by $\rho^*$, the density
     of the charge distribution at rest. This gives a result consistent with (2.28), but
     is evidently wrong, since charge elements are multiply counted during the passage of the light front.
     For example, a contribution to the integral is assigned proportional to the area of
    the cross-hatched region to the left of LF in Fig.2c for $t \le T/5$.
    However, inspection of Fig.2b, showing the actual geometrical
    configuration at $t = T/5$, shows that, because of the parallel
    motion of the charge distribution, LF has crossed only the fraction of the region
    in Fig.2c that is both shaded and cross-hatched, not the entire cross-hatched region.
   In fact, careful inspection of Fig21-6(c) of Ref.~\cite{FeynLW} shows clearly that
   the contribution due to the passage of the light front over the first bin is overestimated. Only the region
   of the charge distribution to the left of the light front as shown in this figure has been sampled
   at this time, not the filled first bin of Fig21-6(b) of Ref.~\cite{FeynLW}. 

\SECTION{\bf{`Relativistic' derivations of the Li\'{e}nard-Wiechert potentials and the
 electromagnetic fields of a uniformly moving charge}}
 As well as the derivation of the LW potentials by consideration of retardation effects,
 as in the original papers of  Li\'{e}nard and Wiechert, text books on classical
 electromagnetism contain alternative derivations, where no retardation effects are considered,
 but instead a relativistic `length contraction' effect is invoked.
  For example, in Ref.~\cite{LLLW1}, the temporal component $\Ai_0$ of the 4-vector electromagnetic
  potential is obtained by Lorentz-transformation into the frame S, where $\Ai_0$ is defined,
  from the frame S$^*$ in which the point-like source charge $Q$ is at rest:
\begin{equation}
 \Ai_0 = \gamma_u \Ai_0^* = \gamma_u \frac{Q}{r^*}  
\end{equation}
 where
 \begin{equation}
 r \equiv |\vec{x}_q-\vec{x}_Q|,~~~r^* \equiv |\vec{x}_q^*-\vec{x}_Q^*|. 
\end{equation}
 The vectors $\vec{x}_q$, $\vec{x}_Q$ ($\vec{x}_q^*$,$\vec{x}_Q^*$) give the position of
  the field point and the source charge, respectively, in the frames S (S$^*$). These coordinates
  are specified at a fixed time in the frame S ---no retardation effects are considered. 
   It is then assumed that the $x$-coordinate separations in the frames S and S$^*$ are related
   by the relativistic length contraction relation:
  \begin{equation}
   x_q^*- x_Q^* = \frac{x_q- x_Q}{\sqrt{1-\frac{u^2}{c^2}}} 
 \end{equation}
  while the $y$ and $z$ separations are the same in both frames.
 It then follows from (4.2) and (4.3) that 
 \begin{equation}
  (r^*)^2 = \frac{(x_q- x_Q)^2+(1-\frac{u^2}{c^2})[(y_q- y_Q)^2+(z_q- z_Q)^2]}{1-\frac{u^2}{c^2}}.
 \end{equation}
 Denoting by $\psi$ the angle between the vectors $\vec{x}_q-\vec{x}_Q$ and $\vec{u}$, (4.4)
 may be written as:
\begin{eqnarray}
  (r^*)^2 & = & r^2 \frac{[\cos^2 \psi+(1-\frac{u^2}{c^2})\sin^2 \psi]}{1-\frac{u^2}{c^2}}
       \nonumber \\
 & = & r^2 \frac{[1-\beta_u^2\sin^2 \psi]}{1-\frac{u^2}{c^2}}.
 \end{eqnarray}
Substituting $r^*$ from (4.5) in (4.1) than gives
 \begin{equation}
 \Ai_0 \equiv \Ai_0({\rm LW})^{PT} = \frac{Q}{r(1-\beta_u^2\sin^2 \psi )^{\frac{1}{2}}}.
 \end{equation}
 This is the `present time' (PT) formula~\cite{PPPT} for the temporal component of the
 retarded LW potential $\Ai_0({\rm LW})^{ret}$ given in Eq.(2.28) above. All quantities
 in (4.6) are defined at the instant that the potential is specified.
 The `present time' form of the 3-vector potential $\vec{\Ai}$ is calculated, in a similar
 manner, to obtain
 \begin{equation}
\vec{\Ai} \equiv \vec{\Ai}({\rm LW})^{PT} = \frac{Q \vec{\beta}_u}{r(1-\beta_u^2\sin^2 \psi)^{\frac{1}{2}}}
 \end{equation} 
  It is interesting to note that that the $\gamma_u$ factor in (4.1), manifesting the
  4-vector character of $\Ai$, is cancelled by a similar factor originating in the
   `length contraction' effect of Eq.(4.3). 
  \par A similar derivation of $\Ai_0({\rm LW})^{PT}$ may be found in Ref.~\cite{PPLW1} where it
   is noted that the change of variables
   \begin{equation}
   x_q^* = \frac{x_q}{\sqrt{1-\frac{u^2}{c^2}}},~~~ y_q^* =  y_q,~~~ z_q^* =  z_q 
 \end{equation} 
 transforms the d'Alembert equation (2.6) into a Poisson equation, the solution of which
 is the Coulomb electrostatic potential $Q/r^*$. Expressing $r^*$ in terms of
  ($x_q$,$y_q$,$z_q$), neglecting a multiplicative factor $\gamma_u$ (which was cancelled
  in the derivation of Ref.~\cite{LLLW1} by the similar factor in the numerator of the right
  side of (4.1)) the potential $\Ai_0({\rm LW})^{PT}$ is obtained. It is only mentioned at the end
  of the calculation that the purely mathematical transformations of Eqs.(4.8) should be interpreted
  as physical transformations predicted by the Lorentz transformation, Unlike in Ref~\cite{LLLW1}
   the scalar and vector potentials are treated in non-relativistic manner, necessitating
   a (tacit) neglect of a factor $\gamma_u$ in order to recover the LW result.
   \par A `relativistic' derivation of the `present time'  formulae for the electric and magnetic
   fields of a uniformly moving charge, by use of a similar `length contraction' ansatz
  as in Refs.~\cite{LLLW1,PPLW1} is found in Jackson's book~\cite{JackLW1}\footnote{A similar derivation is
   found in the widely-used text book on Electricity and Magnetism by Purcell~\cite{Purcell}.},
   The conventional transformation laws of electric and magnetic fields between the frames
   S$^*$ and S;
  \begin{equation}
   \Ei_x = \Ei_x^*,~~~\Ei_y = \gamma_u( \Ei_y^*+\beta_u \Bi_z^*),~~~\Bi_z = \gamma_u( \Bi_z^*+\beta_u \Ei_y^*)
   \end{equation} 
  are used to transform the fields in the rest frame of the source charge:
 \begin{equation}
  \Ei_x^* = \frac{Q(x_q^*- x_Q^*)}{(r^*)^3},~~~\Ei_y^* = \frac{Q(y_q^*- y_Q^*)}{(r^*)^3},~~~\Ei_z^* 
   = \Bi_x^*= \Bi_y^*= \Bi_z^*=0
  \end{equation} 
  into the frame S. Performing this transformation, and using (4.3) to express the result
 in terms of S frame coordinates\footnote{Actually Jackson used a relativistic time dilatation
  equation equivalent to Eq.(4.3).} gives
 \begin{eqnarray}
 \Ei_x & = & \frac{Q(x_q- x_Q)}{\gamma_u^2
 \{(x_q- x_Q)^2+(1-\frac{u^2}{c^2})[(y_q- y_Q)^2+(z_q- z_Q)^2]\}^{\frac{3}{2}}} \nonumber \\
 & = & \frac{Q \cos \psi}{\gamma_u^2 r^2(1-\beta_u^2\sin^2 \psi)^{\frac{3}{2}}}, \\
  \Ei_y & = & \frac{Q(y_q- y_Q)}{\gamma_u^2
 \{(x_q- x_Q)^2+(1-\frac{u^2}{c^2})[(y_q- y_Q)^2+(z_q- z_Q)^2]\}^{\frac{3}{2}}} \nonumber \\
 & = & \frac{Q \sin \psi}{\gamma_u^2 r^2(1-\beta_u^2\sin^2 \psi)^{\frac{3}{2}}}, \\
   \Bi_y & = & \beta_u  E_y =  \frac{Q \beta_u \sin \psi}{\gamma_u^2 r^2(1-\beta_u^2\sin^2 \psi)^{\frac{3}{2}}}.
  \end{eqnarray}
 Eqs.(4.11)-(4.13) may also be written in  3-vector notation as:
  \begin{eqnarray}
 \vec{\Ei} & \equiv &  \vec{\Ei}({\rm H})^{PT} = \frac{Q \vec{r}}{\gamma_u^2 r^3(1-\beta_u^2\sin^2 \psi)^{\frac{3}{2}}}, \\
   \vec{\Bi} & \equiv  &  \vec{\Bi}({\rm H} )^{PT} = \vec{\beta_u} \times \vec{\Ei}.
   \end{eqnarray} 
 The label `H' stands for `Heaviside' who first obtained these equations~\cite{Heaviside}
 more than a decade before the advent of special relativity. They may also be obtained from 
 the `present time' potentials in (4.6) and (4.7) and the usual definitions of electric and magnetic
 fields in terms of derivatives of the 4-vector potential. 
 \par It is easy to show that the `length contraction' ansatz of Eqs.(4.3) and (4.8) used to
  derive (4.14) and (4.15), as obtained from the retarded LW potential, but without invoking
  any retardation effect, is inconsistent with a fundamental reciprocity property of
   special relativity. This was stated in a concise way, and in a manner directly applicable
  to the problem considered here, by Pauli~\cite{Pauli}:
   \par {\tt The contraction of lengths at rest in S$^*$ is equal to that of lengths at\newline
  rest in S and observed in  S$^*$.}
 \begin{figure}[htbp]
\begin{center}
\hspace*{-0.5cm}\mbox{
\epsfysize9.0cm\epsffile{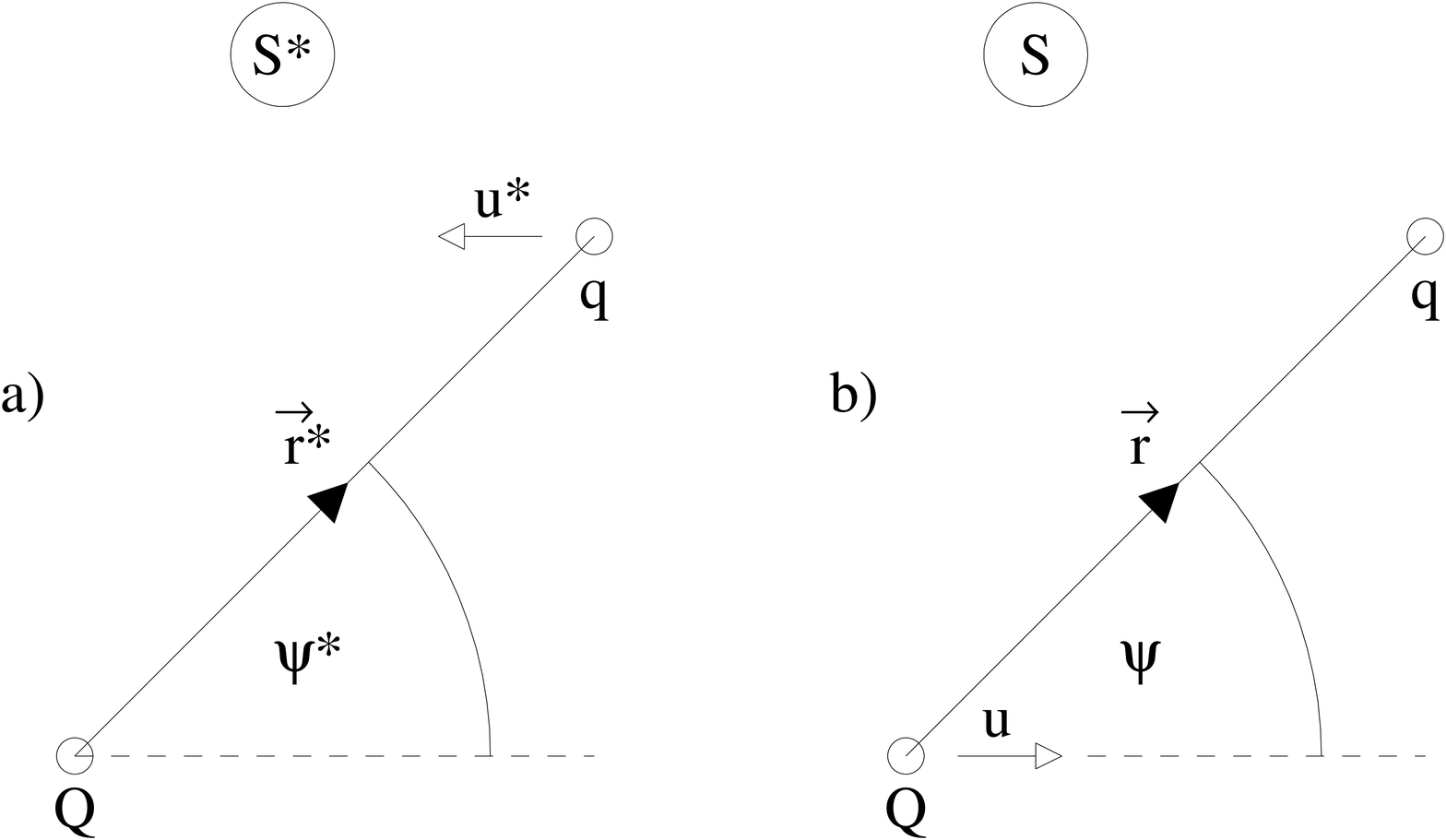}}   
\caption{{\sl Spatial configurations in the frames S$^*$ [a)] and S  [b)] at 
   corresponding instants in the two frames; for example when the origin of  S$^*$, situated
   at $Q$ coincides with the origin of S.}}
\label{fig-fig3}
\end{center}
\end{figure}
 \par To make manifest the symmetry of the configurations in the frames S and  S$^*$, 
   that is the basis of the applicability of the above reciprocity postulate in the present
    case, a test charge $q$, at rest, is
   placed at the field point in S. As shown in Fig.3, the `length at rest in  S$^*$' is
   the separation, $r^*$, of the source and test charges in this frame (Fig.3a). 
     Similarly the  `length at rest in  S is equal to $r$ (Fig.3b).
      However, in the case of the `length contraction' ansatz of Eqs.(4.3) and (4.8), $r$ 
     is also the contracted value of the length $r^*$ as observed in S. i.e.
     \begin{equation}
        r = \alpha(u)r^* 
      \end{equation}
     where $\alpha(u)$ is some even function of the relative velocity $u$ of the frames
     S and S$^*$, and $\alpha(u) < 1$, $\alpha(0) =  1$.
     The above reciprocity postulate states that also
    \begin{equation}
        r^* = \alpha(u^*)r. 
      \end{equation}
        where $\alpha(u) = \alpha(u^*)$
    Combining (4.16) and (4.17),
      \begin{equation}
        r = \alpha(u)r^* =  \alpha(u)\alpha(u^*) r.  
      \end{equation} 
      It follows that if $r \ne 0$,  $\alpha(u)\alpha(u^*) = 1$. This requires that $u = u^*= 0$,
     contradicting the initial hypothesis that S and S$^*$ are in
     relative motion. The existence of a `length contraction' effect respecting Pauli's reciprocity
     condition is therefore excluded by {\it reductio ad absurdum} (self-contradiction). 
     \par The length contraction ansatz of (4.3) and (4.8) is therefore incompatible with
      the above stated reciprocity property of special relativity. How this universally
     (until now) accepted length contraction effect results from a misinterpretation of the symbols
      in the space-time Lorentz transformation has been extensively discussed elsewhere
      ~\cite{JHFSR,JHFSR1,JHFSR2}. In conclusion, the 'relativistic' derivation of the field
    equations (4.14) and (4.15) neglects retardation effects and is in fact incompatible with
     (correctly interpreted~\cite{JHFSR,JHFSR1,JHFSR2})
    special relativity.
     That the same result is obtained using the incorrect
    LW potentials (derived by consideration of pre-relativistic retarded fields)  must then be regarded, not as confirmation of the correctness of
    the formulae, but as purely fortuitous. The `present time' formulae
    derived from the relativistically-correct retarded potentials in (2.12) are 
    presented in the following section.
     \par An alternative `relativistic derivation' of $ \Ai({\rm LW})^{PT}$, but also
   assuming retardation, was given 
     by Landau and Lifshitz~\cite{LLLW2}. The retardation condition (2.11) was used to
      write the temporal component of $\Ai$, in the rest frame of the point-like source
     charge as:
    \begin{equation}
      \Ai_0^* = \frac{Q}{c(t-t'_Q)}.
     \end{equation}
   It was the noticed that the 4-vector:
  \begin{equation}
     \Ai \equiv \frac{Q u}{x^{ret} \cdot u}
     \end{equation}
   where
   \begin{equation}
    x^{ret}  \equiv (c(t-t'_Q);\vec{x}_q-\vec{x}_Q(t'))    
 \end{equation}
   reduces to (4.19) in the rest frame of the source charge. The right side of (4.20)
   is precisely the retarded LW potential in (2.28) of a point-like charge.
   Although it is true that (4.20) gives (4.19) in the rest frame of the source
   charge, the same is true of the different 4-vector potential in (2.12).
   The relation (4.20) is, however, nothing more than a mathematical curiosity,
  lacking any physical motivation, whereas the potential in (2.12), equally
  consistent with (4.19), is the solution of the d'Alembert equations (derived
   from Maxwell's equations and the Lorenz condition) for a point-like charge.
    The physical meaning and method of derivation of the potential of (2.12),
    unlike that of (4.20), are therefore quite clear. 
    \par That the retarded LW potentials and the associated fields could be derived
    in a `relativistic' calculation in which retardation effects are completely 
     neglected, whereas in the original derivations of Li\'{e}nard, Wiechert and
     Heaviside, performed before the advent of special relativity, the (actually
     spurious) length contraction effect is neglected  should be serious cause for concern.
     This unease, however, seemed not to
      be shared by authors of text books, and the pedagogical literature, on classical
      electromgnetism, throughout the last century. There is now ample experimental verification of
      the predictions of correctly interpreted~\cite{JHFSR,JHFSR1,JHFSR2}
      special 
      relativity, and that retardation effects do occur in processes where
      real photons are radiated, so that the corresponding classical fields
      must also be retarded. The contradiction posed by the absence of one or
      the other of two essential, but different, physical phenomena in the two
       different derivations of the Heaviside formulae was clearly stated
     by Jefimenko~\cite{JEFrr}, but the obvious doubt shed by this on the correctness
     of the formulae and/or the derivations, was passed over in silence. In fact, as
      demonstrated in the present paper both the original 19th century and the 
      20th century `relativistic' derivations' are wrong. The former because the
       variation of the effective charge density of the moving charge distribution
       was not taken into account, the latter because the `length contraction'
        effect on which they are based, does not exist. It is proposed in 
      the present paper that the correct relativistic
      retarded potentials of a point-like charge are those  given above in Eq.(2.12).
       The corresponding electric and magnetic fields, for the case of a
      charge in uniform motion, are derived in the following section.
       In Section 6 the retarded fields of accelerated charges are 
       considered, and compared with those derived from the LW potentials
       as well as the well-known formulae of Feynman 
        and Jefimenko as well as some others that have appeared in text books and the pedagogical 
        literature.
        \par The RCED 4-vector potential and current differ from those of CEM by the multiplicative
          factor $\gamma_u$ (see Eq.(2.28) above). This leads to to a breakdown of the Gauss law
          for the electric field of a moving charge ~\cite{JHFRSKO,JHFFT} and covariance-breaking terms in the 
          electrodynamic Maxwell equation (Amp\`{e}re's law)~\cite{JHFFT}. In text books on CEM, the
         validity of the electric field Gauss law for both static and moving source 
         charges is justified by invoking the relativistic length contraction effect of
         Eq.(4.3)~\cite{PPLC}. If the charge density in a moving frame transforms as
        the temporal component of a 4-vector $\propto \gamma_u$, and a volume element
       transforms  $\propto (\gamma_u)^{-1}$ due to relativistic length contraction,
        then the charge within the volume element is Lorentz invariant. Since however,
        as demonstrated above, the length contraction effect is spurious, the effective
        charge actually varies as $\gamma_u$ or as $(u/c)^2$ for small $u$. This effect has
        been experimentally observed in the vicinity of an electrically neutral 
        superconducting magnet~\cite{Edwards}.
 
\SECTION{\bf{Retarded electric and magnetic fields of a point-like charge in uniform
   motion}}
  The electric and magnetic fields corresponding to the 4-vector potential (2.12) are obtained
 by straightforward application of the definitions of electric and magnetic fields in
 terms of derivatives of the potential:
 \begin{eqnarray}
  \vec{\Ei} & \equiv & -\vec{\nabla} \Ai_0- \frac{1}{c}\frac{\partial \vec{\Ai}}{\partial t}, \\
 \vec{\Bi} & \equiv & \vec{\nabla} \times \vec{\Ai}
 \end{eqnarray}
  where, without loss of generality, it may be assumed that the electric field is confined to the $x$-$y$ plane,
  \[  \vec{\nabla} \equiv  \hat{\imath}\frac {\partial ~}{\partial x_q}
   + \hat{\jmath}\frac {\partial ~}{\partial y_q}. \]
  Unit vectors along the $x$- $y$- and $z$-axes are denoted as $\hat{\imath}$,  $\hat{\jmath}$ and
   $\hat{k}$.
  To perform the calculation, the retardation condition
  \begin{equation}
   t' = t-\frac{|\vec{x}_q-\vec{x}_Q(t')|}{c} = t-\frac{r'}{c}   
  \end{equation} 
    must be used to express the derivatives with respect to $t$ in (5.1) in
    terms of $t'$, since the retarded position of the source charge is a function
   of $t'$, not of $t$.
    Assuming that $u$ is constant, (2.12) gives:
 \begin{equation}
   \left. \frac{\partial \Ai_0}{\partial x_q}\right|_{t} = -\frac{Q \gamma_u}{(r')^2}      
 \left. \frac{\partial r'}{\partial x_q}\right|_{t}.    
    \end{equation}
   Differentiating the geometrical relation:
 \begin{equation}
 (r')^2 = [x_q-x_Q(t')]^2 + y_q^2
     \end{equation}
  with respect to $x_q$ gives
  \begin{equation}
r'\left. \frac{\partial r'}{\partial x_q}\right|_{t} =  (x_q-x_Q(t'))\left(1-
    \frac{d x_Q(t')}{d t'} \left. \frac{\partial t'}{\partial x_q}\right|_{t}\right).
  \end{equation}
 Differentiating (5.3) with respect to $x_q$,
   \begin{equation}
 \left.\frac{\partial t'}{\partial x_q}\right|_{t} =
  -\frac{1}{c}  \left.\frac{\partial r'}{\partial x_q}\right|_{t}.
 \end{equation}
 Combining (5.6) and (5.7), rearranging, and noting that $d x_Q(t')/dt' = c \beta_u$ gives
   \begin{equation}
 \left. \frac{\partial r'}{\partial x_q}\right|_{t}
 = \frac{x_q-x_Q(t')}{r'(1-\hat{r}' \cdot \vec{\beta}_u)}
 \end{equation}
  where $ \hat{r}' \equiv \vec{r}'/r'$.
  Combining (5.4) and (5.8)
   \begin{equation}
 \left. \frac{\partial \Ai_0}{\partial x_q}\right|_{t} =
  -\frac{Q \gamma_u [x_q-x_Q(t')]}{(1-\hat{r}' \cdot \vec{\beta}_u)(r')^3}.
 \end{equation}
 An analogous relation is obtained for $\partial \A_0/ \partial y_q$ so that
   \begin{equation}
 -\vec{\nabla}\Ai_0 =
  \frac{Q \gamma_u \vec{r}'}{(1-\hat{r}'\cdot \vec{\beta}_u)(r')^3}.
 \end{equation} 
  Considering now the second term on the right side of (5.1), (2.12) gives
  \begin{equation}
 - \frac{1}{c}\frac{\partial \vec{\Ai}}{\partial t} = 
  - \frac{\hat{\imath}}{c} \left. \frac{\partial \Ai_x}{\partial t}\right|_{x_q,y_q} =
   -\frac{ Q \gamma_u  \vec{\beta}_u}{(r')^2}
     \left. \frac{\partial r'}{\partial t'}\right|_{x_q,y_q}
        \left. \frac{\partial t'}{\partial t}\right|_{x_q,y_q}.  
 \end{equation}
 Differentiating (5.5) with respect to $t'$:
  \begin{equation}
r'\left. \frac{\partial r'}{\partial t'}\right|_{x_q,y_q} =  -[x_q-x_Q(t')]\frac{d x_Q (t')}{d t'}
  \end{equation}
 or 
  \begin{equation}
 \left. \frac{\partial r'}{\partial t'}\right|_{x_q,y_q} =  -c \hat{r}' \cdot \vec{\beta}_u.
  \end{equation}
 Differentiating (5.3) with respect to $t$:
   \begin{equation}
\left. \frac{\partial t'}{\partial t}\right|_{x_q,y_q} = 1-\frac{1}{c}
    \left. \frac{\partial r'}{\partial t'}\right|_{x_q,y_q} \times 
 \left. \frac{\partial t'}{\partial t}\right|_{x_q,y_q} 
 \end{equation}  
 Combining (5.13) and (5.14) and rearranging
  \begin{equation}
 \left. \frac{\partial t'}{\partial t}\right|_{x_q,y_q}
  = \frac{1}{1-\hat{r}' \cdot \vec{\beta}_u}.
 \end{equation}  
 Combining (5.1),(5.10),(5.11),(5.13) and (5.15) gives, for the retarded RCED electric field:
  \begin{equation}
 \vec{\Ei}({\rm RCED})^{ret} =  \left.\frac{Q  \gamma_u}{(1-\hat{r}'\cdot \vec{\beta}_u)}\left[ 
      \frac{[\vec{r}'-\vec{\beta}_u (\vec{r}'\cdot \vec{\beta}_u)]}{(r')^3}
     \right]\right|_{t' = t'_Q}.
 \end{equation} 
  Since $\vec{\Ai} = \hat{\imath} \Ai$,
  \begin{equation}
    \vec{\nabla} \times \vec{\Ai} = -\hat{k} \left. \frac{\partial \Ai_x}{\partial y_q}\right|_{t}
     =-\hat{k} \frac{Q \gamma_u \beta_u}{(r')^2}
     \left. \frac{\partial r'}{\partial y_q}\right|_{t}.
 \end{equation} 
    Similarly to (5.8)
   \begin{equation}
   \left. \frac{\partial r'}{\partial y_q}\right|_{t} = \frac{y_q}{r'(1-\hat{r}' \cdot \vec{\beta}_u)}
 \end{equation} 
  So that 
   \begin{equation}
 \vec{\Bi}({\rm RCED})^{ret} = \vec{\nabla} \times \vec{\Ai} 
      =  \left.\frac{Q \gamma_u \vec{\beta}_u \times \vec{r}' }
  {(r')^3(1-\hat{r}'\cdot \vec{\beta}_u)}\right|_{t' = t'_Q}
    = \vec{\beta}_u \times \vec{\Ei}({\rm RCED})^{ret}.
 \end{equation} 
    Apart from the retarded time argument and an
   overall factor $1/(1-\hat{r}'\cdot \vec{\beta}_u)$ (the Jacobian of the
     transformation from $t$ to $t'$, see Eq.(5.15)) Eqs.(5.16) and (5.19) are the
     same as the formulae for the instantaneous force fields of RCED~\cite{JHFRCED}.

 \begin{figure}[htbp]
\begin{center}
\hspace*{-0.5cm}\mbox{
\epsfysize9.0cm\epsffile{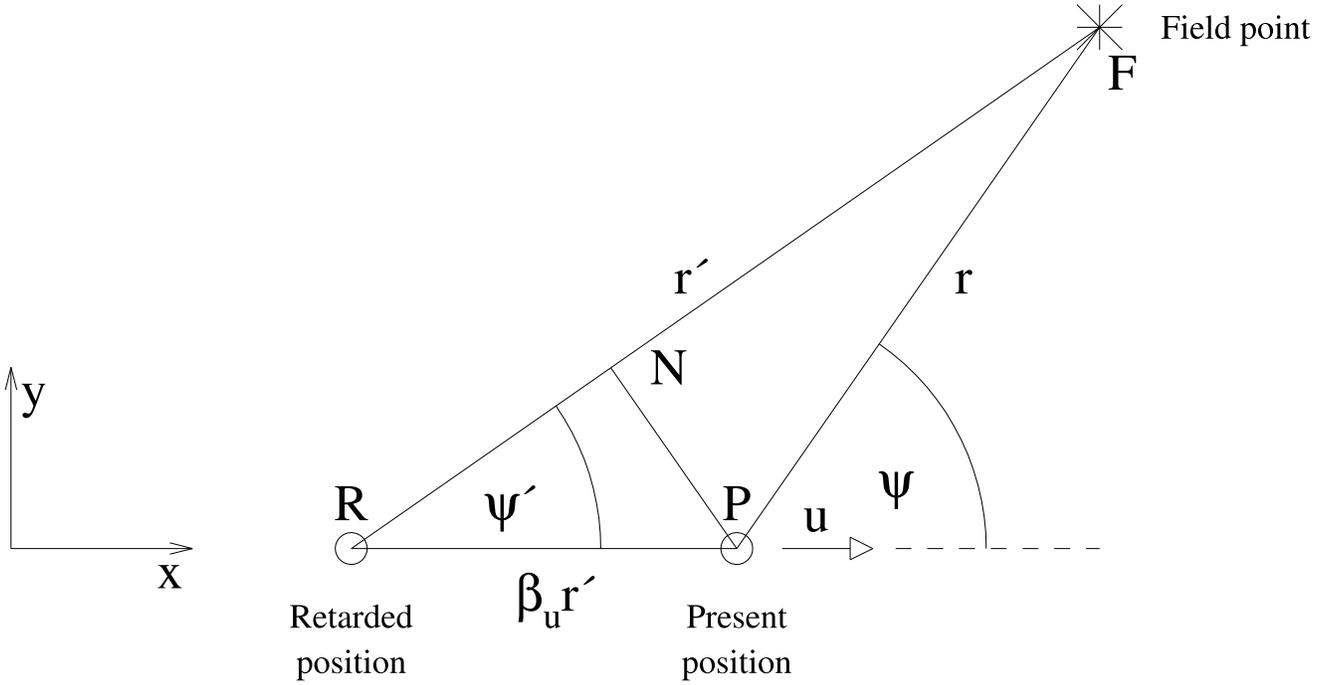}}   
\caption{{\sl Geometry for the calculation of the retarded potential of a charge moving with uniform
 velocity $u$ along the x-axis in terms of the `present time' coordinates $r$, $\psi$. R is the
  position of the charge from which a light signal was emitted so as to arrive at the field
  point F at the instant shown. P is the position of the charge at this instant.
  The line segment PN is perpendicular to RF.}}
\label{fig-fig4}
\end{center}
\end{figure}

    \par For comparison with the Heaviside formulae (4.14) and (4.15), that may be derived from the
      LW potentials it is of interest to write $\vec{\Ei}({\rm RCED})^{ret}$ and $\vec{\Bi}({\rm RCED})^{ret}$
     in the `present time' form.
    \par Consider a point-like charge, $Q$, moving with speed $u$ along the $x$-axis (Fig.4).
     The field point at which the fields are to be specified is denoted by F, the present
    position of the charge by P and the retarded position, lying on the backward light cone 
    of F, by R. If N is the foot of the normal to RF passing through P, the geometry 
    of Fig.4 gives:
    \begin{eqnarray}
      {\rm NF} & = & r'-\beta_u  r'\cos \psi' = r'(1-\hat{r}' \cdot \vec{\beta}_u) \nonumber \\
       & = & \sqrt{r^2-\beta_u^2(r' \sin \psi')^2}
        =\sqrt{r^2-\beta_u^2(r \sin \psi)^2}  \nonumber \\
       &  = & r(1- \beta_u^2 \sin^2 \psi)^{\frac{1}{2}} \equiv r f_u.  
     \end{eqnarray}
         Solving the quadratic equation obtained by applying the cosine rule to the 
       triangle RFP:
    \begin{equation}
         (r')^2 = r^2 + \beta_u^2(r')^2 + 2 \beta_u r r' \cos \psi
   \end{equation} 
       gives the retarded separation between the source charge and field point, $r'$ in terms
    of the 'present time' parameters $r$ and $\psi$
  \begin{equation}
    r' = r\frac{(\beta_u \cos \psi + f_u)}{1-\beta_u^2}.
    \end{equation}  
    Also 
   \begin{equation}
 \sin \psi' = \frac{r \sin \psi}{r'} = \frac{(1-\beta_u^2) \sin\psi}{\beta_u \cos \psi + f_u}
 \end{equation}
   and      
   \begin{eqnarray}
      \cos \psi' & = &  = \frac{\beta_u f_u + \cos \psi}{\beta_u \cos \psi + f_u} \\
      \hat{r}' & = & \hat{\imath} \cos \psi' + \hat{\jmath} \sin \psi'  \\
        \hat{r}' \cdot \vec{\beta}_u & = & \beta_u \cos \psi'.
 \end{eqnarray}
  Eqs.(5.20)-(5.26) may now be used to express the retarded fields in terms of
  `present time' coordinates:
    \begin{eqnarray}
   \vec{\Ei}({\rm RCED})^{PT} & = & \frac{Q (1-\beta_u)[ \hat{\imath}( \beta_u f_u + \cos \psi)
    + \hat{\jmath}(1+\beta_u) \sin \psi]}{r^2 \gamma_u(\beta_u \cos \psi + f_u)^2 f_u}, \\
  \vec{\Bi}({\rm RCED})^{PT} & = & \frac{Q \hat{k} \sin \psi}{r^2 \gamma_u^3(\beta_u \cos \psi + f_u)^2 f_u}
      = \vec{\beta}_u \times \vec{\Ei}({\rm RCED})^{PT}.
\end{eqnarray}

 \begin{figure}[htbp]
\begin{center}
\hspace*{-0.5cm}\mbox{
\epsfysize14.0cm\epsffile{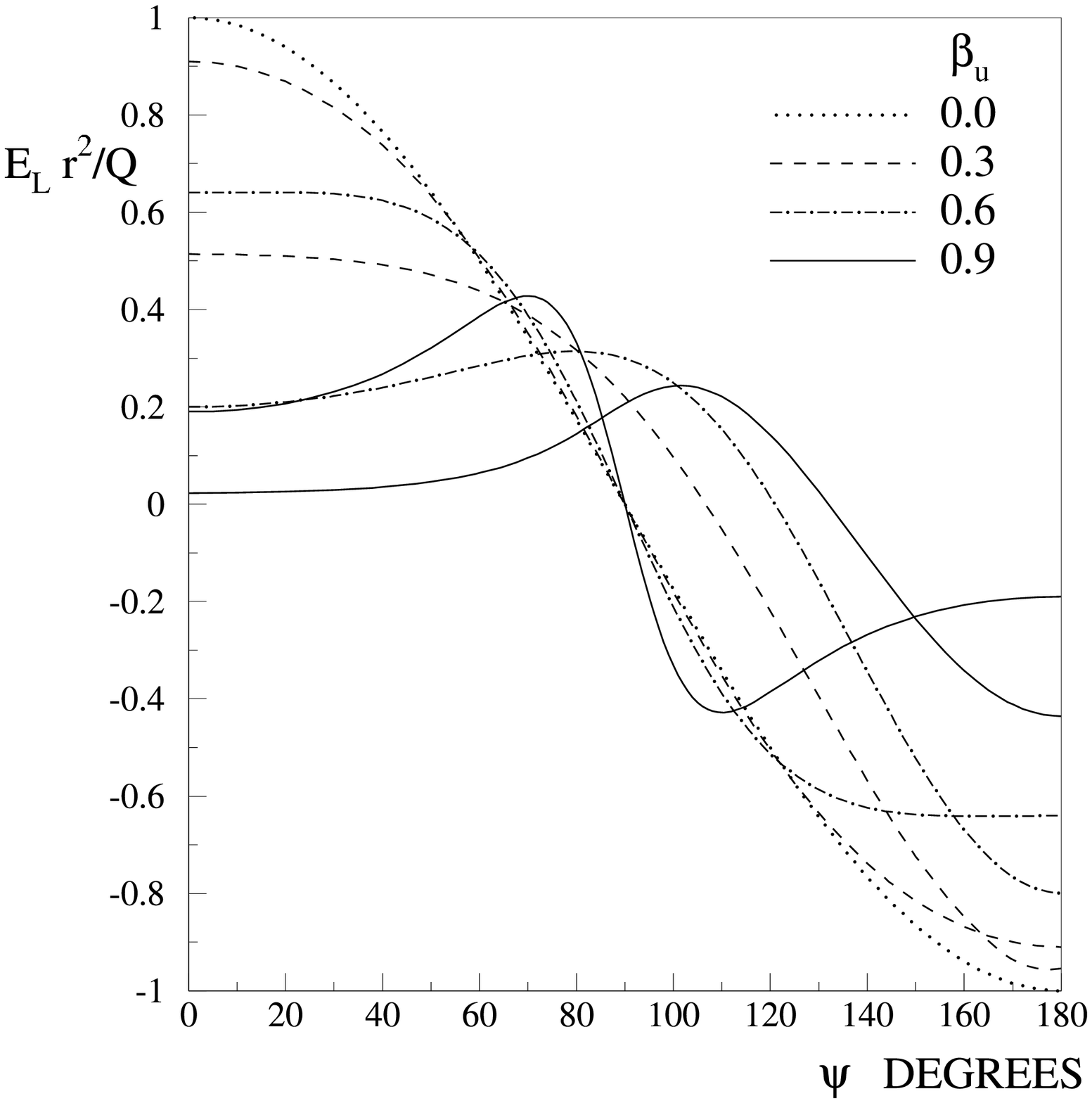}}   
\caption{{\sl Scaled retarded present time longitudinal electric field: $E_Lr^2/Q$ as a function of $\psi$ for various values of the source charge
   velocity $\beta_u = u/c$. The curves of $E_L(H)r^2 /Q$ are antisymmetric relative to
   $\psi = 90^{\circ}$ and so do not display the expected $\psi$ dependence of
   retarded fields as seen in the $E_L(RCED)r^2/Q$ curves where $|E_L(RCED)|$ for $\psi <  90^{\circ}$
   (source charge approaching the field point) is less than $|E_L(RCED)|$ for $\psi >  90^{\circ}$
   (source charge receding from the field point).}}
\label{fig-fig5}
\end{center}
\end{figure}

 \begin{figure}[htbp]
\begin{center}
\hspace*{-0.5cm}\mbox{
\epsfysize14.0cm\epsffile{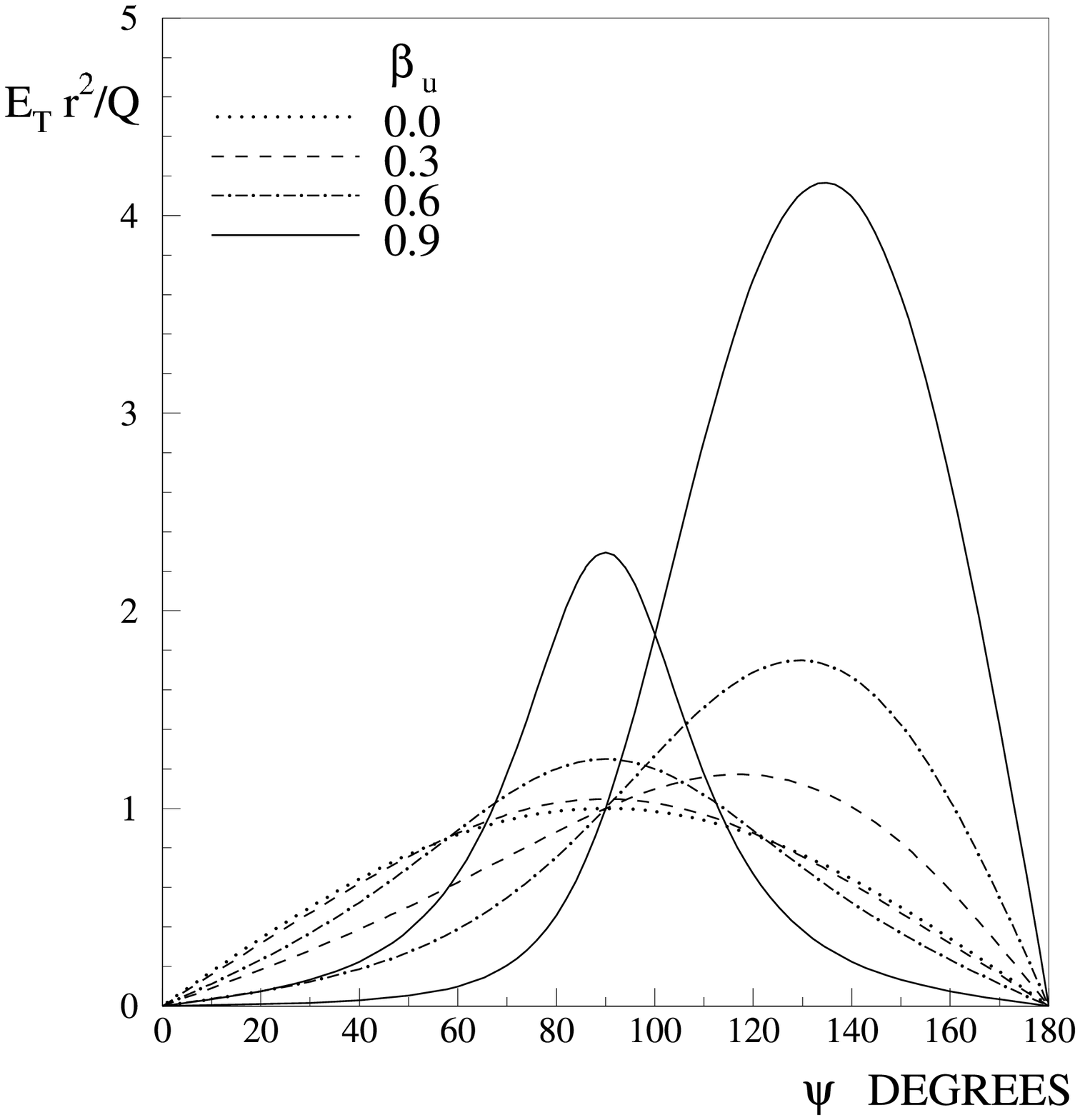}}   
\caption{{\sl  Scaled  retarded present time transverse electric field: $E_Tr^2/Q$ as a function of $\psi$ for various values of the source charge
   velocity $\beta_u = u/c$. The curves of $E_T(H)r^2/Q$ are symmetric relative to
   $\psi = 90^{\circ}$ and so do not display the expected $\psi$ dependence of
   retarded fields as seen in the $E_T(RCED)r^2/Q$ curves where $|E_T(RCED)|$ for $\psi <  90^{\circ}$
   (source charge approaching the field point) is less than $|E_T(RCED)|$ for $\psi >  90^{\circ}$
   (source charge receding from the field point).}}
\label{fig-fig6}
\end{center}
\end{figure}

    These expressions replace, in relativistic classical electrodynamics, the pre-relativistic 
    Heaviside formulae (4.14) and (4.15). Important differences are that (5.27), unlike (4.14),
    is not radial, and in consequence does not respect Gauss' Law, and 
    does not revert to a radial Coulomb field on neglecting terms of O($\beta_u^2$). 
   \par The manifestly incorrect physical behaviour of the retarded electric field given by the
     Heaviside formula (4.14) is evident on comparing it with that given by (5.27). This is done
    in Figs. 5 and 6 which show curves of $\Ei_L^{PT}r^2/Q$ and  $\Ei_T^{PT}r^2/Q$, respectively, as a function
     of $\psi$, where $\vec{\Ei}^{PT} = \hat{\imath}\Ei_L^{PT}+\hat{\jmath}\Ei_T^{PT}$, for different values
     of $\beta_u$, as given by (4.14) and (5.27). Elementary physical considerations require that 
      if $\psi < 90^{\circ}$ (i.e. the source charge is approaching the field point) the magnitude
     of the retarded field must be less than when  $\psi > 90^{\circ}$ and the charge is receding from
     the field point. This is because in the former case the source charge was further from the field
       point than its present-time position when the retarded field was emitted, and closer to it
      in the latter case. Because the strength of the field is inversely proportional to 
      the square of the source-field point separation, at the time of emission of the retarded
      field, the magnitude of the field  must be greater at an angle $\psi_+ =  90^{\circ} + \alpha$ than
      at an angle $\psi_- =  90^{\circ} - \alpha$ where $\alpha > 0$. The fields given by (5.27) 
      demonstrate this property, whereas $\vec{\Ei}({\rm H})^{PT}$ given by (4.14) 
       gives symmetric behaviour for $\Ei_T$:
        \begin{equation}
        \Ei_T({\rm H})^{PT}(\psi_+) = \Ei_T({\rm H})^{PT}(\psi_-)
         \end{equation}
       and antisymmetric behaviour for  $\Ei_L$:
     \begin{equation}
        \Ei_L({\rm H})^{PT}(\psi_+) = -\Ei_L({\rm H})^{PT}(\psi_-).
         \end{equation}
       These symmetry properties are those of instantaneous~\cite{JHFRCED,JHFRSKO}, not retarded,
       force fields\footnote{See, for example, the comparision of the `present time' retarded LW
       fields with the instantaneous RCED fields in Figs. 2 and 3 of Ref.~\cite{JHFRSKO}.}. As explained in
       Section 2 above, the antisymmetry of the $E_L({\rm H})^{PT}$ curves in Fig. 5, about
        $\psi = 90^{\circ}$ and the symmetry of the $E_T({\rm H})^{PT}$ curves in Fig. 6, about
        $\psi = 90^{\circ}$ in Fig. 6 is a consequence of the neglect of the velocity dependence
        of the source charge density in deriving the LW potentials of Eq.(2.28).
\SECTION{\bf{Retarded electric and magnetic fields of an accelerated point-like charge:
     the RCED, Li\'{e}nard-Wiechert, Feynman and Jefimenko equations}}
     The generalisation of the RCED formulae (5.16) and (5.19) to the case of a non-constant
     value of the source charge velocity $\vec{u}$ is straightforward. The details of the calculation
     may be found in Ref.~\cite{JHFFT}. Including the terms generated by the acceleration of the source
    charge gives:
 \begin{eqnarray}
 \vec{\Ei}({\rm RCED})^{ret} & = & \left. \left\{\frac{Q  \gamma_u}{K}\left[ 
 \frac{[\hat{r}'-\vec{\beta}_u (\hat{r}'\cdot \vec{\beta}_u)]}{(r')^2}
  +\frac{[\gamma_u^2 \beta_u \dot{\beta}_u(\hat{r}'- \vec{\beta}_u)- \dot{\vec{\beta}_u}]}
   {c r'} \right]\right\}\right|_{t' = t'_Q}, \\
 \vec{\Bi}({\rm RCED})^{ret} & = & \left. \left\{ \frac{Q \gamma_u (\vec{\beta}_u \times \hat{r}')}{K}
\left[ \frac{1}{(r')^2} + \frac{\gamma_u^2 \dot{\beta}_u}{c \beta_u r'}  \right]\right\}\right|_{t' = t'_Q}
 \end{eqnarray}
  where
  \begin{equation}
   K \equiv (1- \hat{r}' \cdot \vec{\beta}_u),~~~\dot{\beta}_u \equiv |\dot{\vec{\beta}_u}|.
  \end{equation} 
   It follows from (6.1) that
 \begin{equation}
 \vec{\beta}_u \times \vec{\Ei}({\rm RCED})^{ret}  = \left. \left\{ \frac{Q \gamma_u (\vec{\beta}_u \times \hat{r}')}{K}
\left[ \frac{1}{(r')^2} + \frac{\gamma_u^2\beta_u \dot{\beta}_u}{c r'}  \right]
  -\frac{Q\gamma_u(\vec{\beta}_u \times \dot{\vec{\beta}_u})}{Kcr'} \right\}\right|_{t' = t'_Q}
 \end{equation}
 and 
  \begin{equation}
 \hat{r}'\times \vec{\Ei}({\rm RCED})^{ret} = \left. \left\{ \frac{Q \gamma_u (\vec{\beta}_u \times \hat{r}')}{K}
\left[ \frac{\hat{r}'\cdot \vec{\beta}_u}{(r')^2} + \frac{\gamma_u^2\beta_u \dot{\beta}_u}{c r'}  \right]
  -\frac{Q\gamma_u( \hat{r}' \times \dot{\vec{\beta}_u})}{Kcr'} \right\}\right|_{t' = t'_Q}.
 \end{equation} 
  The relation $\vec{\Bi}({\rm RCED})^{ret} =  \vec{\beta}_u \times \vec{\Ei}({\rm RCED})^{ret}$ than holds only if
   $\dot{\beta}_u =0$, as in Eq.(5.19) above, while, in all cases,
      $\vec{\Bi}({\rm RCED})^{ret} \ne \hat{r}'
  \times \vec{\Ei}({\rm RCED})^{ret}$ 
 
  \par These formulae may be compared with those derived by inserting the LW potentials of Eq.(2.28) into the defining
  equations (5.1) and (5.2) of the electric and magnetic fields, making use of the 
  Jacobian of (5.15) to relate derivatives with respect to $t$ and $t'$. This calculation is given
    in Appendix A. It is found that:
  \begin{eqnarray}
 \vec{\Ei}({\rm LW})^{ret} & = &  \left.\left\{\frac{Q}{K^3}\left[\frac{\hat{r}'- \vec{\beta}_u}{\gamma_u^2 (r')^2}
    + \frac{\hat{r}' \times [(\hat{r}'- \vec{\beta}_u) \times \dot{\vec{\beta}_u}]}{c r'}\
  \right]\right\}\right|_{t' = t'_Q}, \\
 \vec{\Bi}({\rm LW})^{ret} & = & \left. \left\{\frac{Q (\vec{\beta}_u \times \hat{r}')}{K^3}\left[\frac{1}
 {\gamma_u^2(r')^2}  +\frac{(\dot{\beta}_u(1-\hat{r}' \cdot \vec{\beta}_u)
 + \beta_u(\hat{r}' \cdot \dot{\vec{\beta}_u})}{c r' \beta_u}\right]
   \right\} \right|_{t' = t'_Q}. 
 \end{eqnarray}
  The markedly different angular dependence of these fields to that of the RCED fields of (6.1) and (6.2) may be
  noticed.
    Eq.(6.6) gives
  \begin{equation}
  \vec{\beta}_u \times  \vec{\Ei}({\rm LW})^{ret} =  \left. \left\{\frac{Q}{K^3}
   \left[(\vec{\beta}_u \times \hat{r}')\left(\frac{1}{\gamma_u^2(r')^2}
  +\frac{\hat{r}' \cdot \dot{\vec{\beta}_u}}{c r'}\right)
   \right]
   \right\} \right|_{t' = t'_Q},
\end{equation}
 \begin{equation}
   \hat{r}' \times  \vec{\Ei}({\rm LW})^{ret} =  \vec{\Bi}({\rm LW})^{ret}
\end{equation}
 The relation $\vec{\Bi}({\rm LW})^{ret} = \vec{\beta}_u \times  \vec{\Ei}({\rm LW})^{ret}$
  then hold only if  $\dot{\beta}_u = 0$.
 
  \par The RCED retarded fields (6.1) and (6.2) are now compared with those obtained from
   formulae given by Feynman, Jefimenko and other authors.
  The consistency of the latter fields with the LW ones of (6.6) and (6.7) will also be considered.
   In the `Feynman Lectures in Physics' compact formulae for the retarded fields of an accelerated  point-like charge are given, 
  but not derived~\cite{FeynFMC1, FeynFMC2}. In the notation of the present paper they are
   \begin{eqnarray}
  \vec{\Ei}({\rm Feyn})^{ret} & = & Q \left. \left[ \frac{\hat{r}'}{(r')^2}+\frac{r'}{c}\frac{d ~}{d t}\left(
    \frac{\hat{r}'}{(r')^2}\right) + \frac{1}{c^2} \frac{d^2 \hat{r}'}{d t^2} \right]\right|_{t' = t'_Q}, \\
 \vec{\Bi}({\rm Feyn})^{ret} & = &  \left. \hat{r}'\right|_{t' = t'_Q} \times   \vec{\Ei}({\rm Feyn})^{ret}.
 \end{eqnarray} 
 Since (see Eq.(5.3)) $r'$ is a function of the retarded time $t'$, not of the present time $t$ it is necessary
   to introduce the Jacobian of Eq.(5.15) in order to evaluate the derivatives in Eq.(6.10). Although Feynman uses
    the symbol for a total time derivative rather than a partial one in Eq.(6.10) it is clear from
    the definitions of the fields in terms of potentials in (5.1) and (5.2) that the time derivatives
    should be understood as partial ones for a fixed position of the field point $\vec{x}_q$ as in 
    (5.15).
   The straightforward but somewhat lengthy calculation, which is analogous to that shown in the
    previous section, leading
    to Eqs.(5.16) and (5.19), is presented in Appendix B. The following formula for the electric field is
    obtained:
  \begin{eqnarray}
    \vec{\Ei}({\rm Feyn})^{ret} & = & Q \left\{\frac{\hat{r}'}{(r')^2} + \frac{1}{K}
  \frac{[3 \hat{r}'(\hat{r}'\cdot \vec{\beta}_u)-\vec{\beta}_u]}{(r')^2}\right. \nonumber \\
    &   & + \frac{1}{K^2}\left[ \frac{\hat{r}'[3 (\hat{r}'\cdot \vec{\beta}_u)^2 -\beta_u^2] - 2 \vec{\beta}_u
  (\hat{r}'\cdot \vec{\beta}_u)}{(r')^2}+ \frac{[\hat{r}'(\hat{r}' \cdot \dot{\vec{\beta}_u})
   - \dot{\vec{\beta}_u}]}{cr'}\right]
 \nonumber \\
     &   & + \left. \left. \frac{[\hat{r}'(\hat{r}'\cdot \vec{\beta}_u)- \vec{\beta}_u]}{K^3}
       \left[ \frac{(\hat{r}'\cdot \vec{\beta}_u)^2 -\beta_u^2}{(r')^2}
      +\frac{\hat{r}' \cdot \dot{\vec{\beta}_u}}{cr'} \right]\right\}\right|_{t' = t'_Q}.
  \end{eqnarray}
   Collecting terms on the right side of (6.12) proportional to $\hat{r}'$, $\vec{\beta}_u$
   and $\dot{\vec{\beta}_u}$ recovers, together with (6.11), the LW formulas (6.6) and (6.7).

  \par A formula for the retarded potentials, similar to Eq.(2.8) above, is obtained in
    in Ref.~\cite{JPS} by using Green functions to solve the inhomogeneous d'Alembert
     equations (2.6) and (2.7). However, subsequently, the usual mistake is made of
     neglecting the motion-dependence of the charge density, as explained in Section 2.
     After performing the spatial integration, instead of replacing 
     $r'(t')$ in the argument of the $\delta$-function by  $r'(t_ Q')$, (see Eq.(2.10) above) 
     the appropriate retarded position of the point-like source charge at time $t$, the
     same functional dependence on $t'$ is assumed as before the spatial integration
     and the formula (2.18) is then used to transform the argument of the  $\delta$-function,
     leading to the spurious retardation factor $1/K$ of the LW potentials.
      In this way the formula (19) of Ref.~\cite{JPS} was obtained which was then shown to
     give the Feynman's formula Eq.~(6.10) above. It was also correctly stated (but not
     demonstrated) that the same formula was equivalent to the LW field of Eq.~(6.6) above.

     \par The Jefimenko formulae for the fields of an accelerated charge distribution are
      ~\cite{Jefimenko}:
      \begin{eqnarray}
      \vec{\Ei}({\rm Jefi})^{ret} & = & \int \left\{\hat{r}' \left[\frac{[\rho]}{(r')^2}+ \frac{1}{cr'}
       \frac{\partial[\rho]}{\partial t}\right]-\frac{1}{c^2r'} 
       \frac{\partial[\vec{\Jnr}]}{\partial t}\right\} d^3 x_{\Jnr}, \\
  \vec{\Bi}({\rm Jefi})^{ret} & = &  \frac{1}{c}\int \left[\frac{[\vec{\Jnr}]}{(r')^2}+ \frac{1}{cr'}
       \frac{\partial[\vec{\Jnr}]}{\partial t}\right] \times \hat{r}' d^3 x_{\Jnr}.
    \end{eqnarray}
     The square brackets around the charge density $\rho$ and the non-relativistic
     current density $\vec{\Jnr}$ indicate that they are evaluated at the retarded time
    $t' = t-r'/c$, as is also the spatial separation $r'$ of the current element from 
    the field point. The volume element, $d^3 x_{\Jnr}$, is also specified at the 
    retarded time $t'$. Important differences from the RCED, LW and Feynman formulae
     are that the time derivatives act only on the charge and current densities, not on $r'$,
     and that, as compared to the Feynman formula, only first order time derivatives 
     appear. However  a time derivative of $\vec{r}'$ is implicit in the definition
     of the current $\vec{\Jnr}$.
     \par In order to discuss the Jefimenko equations for the case of a point-like charge,
       it will be found convenient to explicitly impose the retardation condition 
       by including an integration over the retarded time $t'$ together with the
       corresponding $\delta$-function as in Eq.(2.8). Indeed, much confusion about,
       and misinterpretation of, formulae for retarded fields result from not properly
        taking into account integrations over both space and time. This must be done to
        correctly describe the essential physical properties of the problem under
      consideration ---a spatially extended distribution of charge\footnote{In the real 
      world, consisting of an ensemble of identical point-like charged particles.} in motion that
      is probed, in time, by the backward light cone of the field point. As will be seen, 
       attempts to simplify formulae by omitting time integrals, 
       and the associated $\delta$-functions, as in (6.13) and (6.14),
       and in many text-book treatments of retarded potentials, often lead to erroneous
       results.
      Specifying explicitly the retardation condition, (6.13) and (6.14) are written as:
    \begin{eqnarray}
     \vec{\Ei}({\rm Jefi})^{ret} & = &  \int dt' \int \left\{ \left[\frac{[\rho(\vec{x}_{\Jnr}(t'),t')}
      {(r')^2}+ \frac{1}{cr'}
       \frac{\partial[\rho(\vec{x}_{\Jnr}(t'),t')]}{\partial t}\right]\hat{r}'\right. \nonumber \\
    &  & \left. -\frac{1}{c^2r'} \frac{\partial[\vec{\Jnr}(\vec{x}_{\Jnr}(t'),t')]}{\partial t}\right\}
       \delta(t' +\frac{r'(t')}{c}-t) d^3 x_{\Jnr}(t'), \\
  \vec{\Bi}({\rm Jefi})^{ret} & = &  \int dt' \int \frac{1}{c} \left[\frac{[\vec{\Jnr}(\vec{x}_{\Jnr}(t'),t')]}{(r')^2}
\right. \nonumber \\
    &  & \left.     + \frac{1}{cr'}
    \frac{\partial[\vec{\Jnr}(\vec{x}_{\Jnr}(t'),t')]}{\partial t}\right] \times \hat{r}'
      \delta(t' +\frac{r'(t')}{c}-t) d^3 x_{\Jnr}(t').
    \end{eqnarray}
     A point-like charge $Q$ has, in non-relativistic approximation, the following
    charge and current densities:
       \begin{eqnarray}
   \rho_Q(\vec{x}_{\Jnr}(t'),t') = Q \delta(\vec{x}_{\Jnr}(t')-\vec{x} _Q(t')), \\
    \vec{\Jnr}_Q(\vec{x}_{\Jnr}(t'),t') = Q \vec{u}(t')\delta(\vec{x}_{\Jnr}(t')-\vec{x} _Q(t')).      
    \end{eqnarray}
    Substituting (6.17) and (6.18) into (6.15) and (6.16) and performing the spatial integrations
     gives:
    \begin{eqnarray}
      \vec{\Ei}({\rm Jefi})^{ret} & = & Q \int dt' \left[\frac{\hat{r}'}
      {(r')^2} -\frac{1}{c^2r'} \frac{\partial \vec{u}(t')}{\partial t}\right]
       \delta(t'-t'_Q) \nonumber \\
 & = &  \left .\left . Q \left[\frac{\hat{r}'}{(r')^2} -\frac{1}{c^2r'}\frac{\partial t'}
  {\partial t}\right|_{\vec{x}_q} 
  \frac{d \vec{u}(t')}{d t'}\right]\right|_{t'= t'_Q}  \nonumber \\
 & = & \left. Q \left[\frac{\hat{r}'}{(r')^2} -\frac{1}{K c^2r'}
  \frac{d\vec{u}(t')}{d t'}\right]\right|_{t' = t'_Q},  \\
  \vec{\Bi}({\rm Jefi})^{ret} & = &  Q \int dt' \left[\frac{\vec{u}(t')}{c (r')^2}
   + \frac{1}{c^2r'} \frac{\partial \vec{u}(t')}{\partial t}\right] \times \hat{r}'
     \delta(t'-t'_Q) \nonumber \\
 & = &  Q\left\{ \left[\frac{\vec{u}(t')}{ c (r')^2}
   \left. \left.+ \frac{1}{c^2r'}\frac{\partial t'}{\partial t}\right|_{\vec{x}_q}
     \frac{d \vec{u}(t')}{d t'}\right] \times \hat{r}'
   \right\} \right|_{t' = t'_Q} \nonumber \\
 & = &  Q\left\{ \left[\frac{\vec{u}(t')}{c (r')^2}
    \left.+ \frac{1}{Kc^2r'}
     \frac{d\vec{u}(t')}{d t'}\right] \times \hat{r}'
    \right\} \right|_{t'= t'_Q}
    \end{eqnarray}
     where
  \begin{equation}
           t'_Q \equiv t - \frac{r'(t'_Q)}{c}.
   \end{equation}
     For a uniformly moving charge, in contrast to the RCED, LW and Feynman equations
      the Jefimenko equations therefore predict the same (Coulombic) electric field as 
      in the electrostatic case.
     \par In a paper on time-dependent generalisations of of the Biot and Savart and Coulomb
       laws where the Jefimenko equations were extensively discussed~\cite{GH},
    it was claimed that the Jefimenko, Li\'{e}nard-Wiechert and Feynman
     formulae for the retarded fields are all consistent with each other. The arguments
     given in support of this conclusion are critically examined below, but first the
     claim of the authors of Ref.~\cite{GH} to {\it derive} the  Jefimenko equations
       from the defining formulae (5.1) and (5.2) of electric and magnetic fields 
       and retarded potentials is considered. The assumed form of the potentials in Ref.~\cite{GH}
      in the notation and choice of units of the present paper, is:
       \begin{eqnarray}
        \Ai_0 & = & \int \frac{[\rho]}{r'}d \tau, \\
   \vec{\Ai} & = & \int \frac{[\vec{\Jnr}]}{r'}d \tau.
  \end{eqnarray}
    The volume element $d \tau \equiv d^3x_{\Jnr}(t')$ and the quantities in square brackets,
     as well as the distance $r'$ between the volume element and the field point
      are all specified at the retarded time $t' = t-r'/c$.
       Note that unlike the solutions of the D'Alembert equations in (2.8) there is no
       integral over the retarded time in these equations and also they do not contain the
      $1/K$ factor of the LW potentials. Substitution of (6.22) and (6.23)
        into (5.1) gives\footnote{The field is labelled according to the initials
         of the authors, Griffiths and Heald, of Ref.~\cite{GH}}:
       \begin{equation}
       \vec{\Ei}({\rm GH1})^{ret} = -\int \left[ \frac{1}{r'}\vec{\nabla}[\rho] +
       [\rho]\vec{\nabla}\left(\frac{1}{r'}\right)+ \frac{1}{c^2 r'}\frac{\partial [\vec{\Jnr}]}
       {\partial t} + \frac{[\vec{\Jnr}]}{c^2}\frac{\partial ~}
       {\partial t}\left(\frac{1}{r'}\right)\right]d \tau.
       \end{equation}
      This aleady disagrees with the corresponding equation (21) of Ref.~\cite{GH}, where
      the last term on the right side of (6.30) is omitted. Indeed, this term does not 
      vanish but the last factor in it is:
        \begin{equation}
   \left.  \frac{\partial ~}{\partial t}\left(\frac{1}{r'}\right)\right|_{\vec{x}_q} =
      \left. \frac{\partial t'}{\partial t}\right|_{\vec{x}_q} \left.\frac{\partial ~}
      {\partial t'}\left(\frac{1}{r'}\right)\right|_{\vec{x}_q} 
       = -\frac{1}{(r')^2} \left. \frac{\partial t'}{\partial t}\right|_{\vec{x}_q}
      \left.\frac{\partial r'}
      {\partial t'}\right|_{\vec{x}_q} = \frac{c(\hat{r'} \cdot \vec{\beta_u})}{K (r')^2} 
   \end{equation}
     where (5.13) and (5.15) have been used. 
    This result is implicit in the later Eq.(44) of Ref.~\cite{GH}, so the omission
    of the term in Eq.(21) of this reference in hard to understand. The retarded density $[\rho]$ 
     may be a function of the retarded time $t'$ and the position $\vec{x}_{\Jnr}(t')$ of the
     volume element $d \tau$, but does not depend of the position $\vec{x}_q$ of the field
    point. Therefore $ \vec{\nabla}[\rho]$ vanishes. More formally\footnote{The ellipsis in (6.26) and
   subsequent equations indicates the contribution of the $x$- and $y$-components.}
   \begin{eqnarray}
     \vec{\nabla}[\rho] & =  &  \left. \hat{\imath} \frac{\partial [\rho]}{\partial x_q}\right|_{t}
        +~.~.~. \nonumber \\
      & =  &  \left. \hat{\imath}  \frac{\partial t'}{\partial x_q}\right|_{t} \frac{d [\rho]}{d t'}
        +~.~.~. \nonumber \\
  & =  &  \left. \left. \left. \hat{\imath}  \frac{\partial t'}{\partial x_q}\right|_{t} \frac{\partial t}
   {\partial t'}\right|_{t} \frac{\partial [\rho]}{\partial t}\right|_{t}  +~.~.~. \nonumber \\
     & =  &  0
  \end{eqnarray}
     to be compared with the relation
       \begin{equation}
     \vec{\nabla}[\rho] = -\frac{1}{c}\frac{\partial [\rho]}{\partial t}\hat{r}'
      \end{equation}
     given in Ref.~\cite{GH}.
     The mathematical error leading to the incorrect equation (6.27) is a subtle
      one concerning the precise definitions of partial derivatives.
      Combining  Eqs.(5.7) and (5.8) gives:
     \begin{equation}
  \left. \frac{\partial t'}{\partial x_q}\right|_{t} = -\frac{1}{c}\frac{(x_q - x_Q)}{K r'}  
    \end{equation}
     so that the second line of (6.26) may be written as:
        \begin{equation}
        \vec{\nabla}[\rho] =  -\frac{\hat{\imath}}{c}\frac{(x_q - x_Q)}{K r'}\frac{d [\rho]}{d t'}  +~.~.~.~~.
         \end{equation}
     Now it seems plausible, in view of Eq.(5.15) above to make the substitution
      \begin{equation}
    \frac{1}{K} \frac {d ~}{d t'} \rightarrow \frac{\partial ~}{\partial t}   
    \end{equation}
     in (6.29) thus yielding (6.27).  But all spatial partial derivatives in (5.1) and
    hence in (6.26) and
     (6.29) are evaluated {\it at constant} $t$ whereas the operator relation of (6.30) is
      (see Eq.(5.15)) valid only {\it at constant} $\vec{x}_q$, and so is inapplicable in relation to
     derivatives with respect to $x_q$, $y_q$ or $z_q$.
     \par In fact the spurious relation (6.27) was also used by
     Jefimenko in the original derivation of Eq.(6.13)~\cite{Jefimenko}.
     Maxwell's equations and Eq.(6.23),
     called the `Vector Identity'
     V-33~\cite{JefiVI}, are used to obtain (6.19) from an integral
     vector identity: the `Vector wave field theorem' V-31~\cite{JefiVI}. 
     \par  The term containing $\vec{\nabla}(1/r')$ in Eq.(6.24) is the same, up to a constant multiplicative factor,
    as one which has been previously calculated in Section 5 (Eq.(5.10)) so that:
    \begin{equation}
    -\vec{\nabla}\left(\frac{1}{r'}\right) = \frac{\hat{r}'}{K(r')^2}.
    \end{equation}
     Combining (6.24), (6.25), (6.26) and (6.27) gives:
    \begin{equation}
       \vec{\Ei}({\rm GH1})^{ret} = \int \left[\frac{[\rho]\hat{r}'-(\hat{r}'\cdot \beta_u)[\vec{\Jnr}]}
      {K(r')^2} - \frac{1}{c^2 r'}\frac{\partial [\vec{\Jnr}]}
       {\partial t}\right]d \tau.
   \end{equation}
      Note that the first term on the right side of (6.32) differs from the corresponding one in
     Jefimenko's formula by a factor $1/K$. This factor was missed in the calculation of
     Ref.~\cite{GH}. In summary, the claimed-to-be-derived Jefimenko formula, Eq.(19) of
     Ref.~\cite{GH} is missing a factor $1/K$ on the first term; the second term vanishes,
      and the fourth term in (6.24) (the second in (6.32)) is omitted. Indeed, only the last term
     of Eq.(19) of Ref.~\cite{GH} is correct. 
      \par Combining (5.2) and (6.23) gives: 
       \begin{equation}
       \vec{\Bi}({\rm GH1})^{ret} = \frac{1}{c}\int \left[\vec{\nabla} \times \frac{[\vec{\Jnr}]}{r'}
       - [\vec{\Jnr}] \times \vec{\nabla}\left(\frac{1}{r'}\right)\right]d \tau.
       \end{equation}
 Choosing $[\vec{\Jnr}]$ parallel to the $x$-axis
      \begin{equation}
      \vec{\nabla} \times [\vec{\Jnr}] = -\hat{k}\left. \frac{\partial |\vec{\Jnr}|}{\partial y_q}
       \right |_t =  -\hat{k}\left.\left.\left. \frac{\partial t'}{\partial y_q} \right |_t
      \frac{\partial t}{\partial t'} \right |_t
        \frac{\partial |\vec{\Jnr}|}{\partial t}\right |_t = 0
 \end{equation}
       whereas the authors of  Ref.~\cite{GH} state that
    \begin{equation}
      \vec{\nabla} \times [\vec{\Jnr}] = \frac{1}{c}\frac{\partial [\vec{\Jnr}]}{\partial t}
       \times \hat{r}'.
 \end{equation}
     This results from a similar misuse of partial derivatives to that described above
     in connection with Eq.(6.27). 
   Eqs.(6.33),(6.31) and (6.34) give
 \begin{equation}
       \vec{\Bi}({\rm GH1})^{ret} = \int \frac{[\vec{\Jnr}] \times \hat{r}'}{c K (r')^2} d \tau
       \end{equation}
   which differs from the Jefimenko equation (6.15) by an overall factor $1/K$ and the absence
 of the $\partial [\vec{\Jnr}]/\partial t$ term. Again the `derivation' of the Jefimenko equation
 in Ref.~\cite{GH} , based now on the incorrect formula (6.35), is
 fallacious. Jefimenko~\cite{Jefimenko} also assumed this formula in
 order to derive the second term on the right side of (6.14).
   \par In Section IV of Ref.~\cite{GH} it is claimed to derive the LW fields of Eqs.(6.4) and (6.5)
 from the Jefimenko formulae. However this derivation starts not from the Jefimenko formulae
 (6.13) and (6.14) but instead from the different equations~\footnote{Eq.~(6.37)is Eq.~(38) of Ref.~\cite{GH}}
 (given here the label `GH2'):
       \begin{eqnarray}
      \vec{\Ei}({\rm GH2})^{ret} & = & \int  \left[\frac{[\rho]\hat{r}'}{(r')^2}+
       \frac{\partial ~}{\partial t}\left(\frac{[\rho]\hat{r}'}{cr'}\right)
       -  \frac{\partial ~}{\partial t}\left(\frac{[\vec{\Jnr}]}{c^2r'}\right)\right] d^3 x_{\Jnr}, \\
  \vec{\Bi}({\rm GH2})^{ret} & = & \int \left[\frac{[\vec{\Jnr}] \times 
   \hat{r}'}{c(r')^2}+
       \frac{\partial ~}{\partial t}\left(\frac{[\vec{\Jnr}] \times 
   \hat{r}'}{c^2r'}\right)\right] d^3 x_{\Jnr}
    \end{eqnarray}
    which differ from the Jefimenko equations in that the time derivatives operate not only
   on the charge and current densities but also on the reciprocal of the retarded
   source-field point separation $r'$ and the unit vector $\hat{r}'$.
   \par The authors of Ref.~\cite{GH} introduce into Eqs.(6.37) and (6.38) point-like
  non-relativistic charge and current densities according to Eq.(6.17) and (6.18).
  Since the integration over $t'$ is omitted, it is then implicit in these equations
   that $t' = t'_Q$, where the fixed time, $t'_Q$, is the solution of Eq.(6.21), 
    corresponding to a fixed position of the source charge for given values of $\vec{x}_q $ and $t$.
  It then follows that for point-like charges (6.37) and (6.38) are written as:
       \begin{eqnarray}
      \vec{\Ei}({\rm GH2})^{ret} & = & Q \int  \left[\frac{\hat{r}'}{(r')^2}+
       \frac{\partial ~}{\partial t}\left(\frac{\hat{r}'}{cr'}\right)
       -  \frac{\partial ~}{\partial t}\left(\frac{\vec{\beta}_u}{c r'}\right)\right]
     \delta(\vec{x}_{\Jnr}(t'_Q)-\vec{x}_Q(t'_Q)) d^3 x_{\Jnr} \nonumber \\
 & = & \left.Q \left[\frac{\hat{r}'}{(r')^2}+
       \frac{\partial ~}{\partial t}\left(\frac{\hat{r}'}{cr'}\right)
       -  \frac{\partial ~}{\partial t}\left(\frac{[\vec{\beta}_u]}{c r'}\right)\right]
     \right|_{t' = t'_Q},  \\
  \vec{\Bi}({\rm GH2})^{ret} & = & Q \int \left[\frac{\vec{\beta}_u \times 
   \hat{r}'}{(r')^2}+
       \frac{\partial ~}{\partial t}\left(\frac{\vec{\beta}_u \times 
   \hat{r}'}{c r'}\right)\right] \delta(\vec{x}_{\Jnr}(t'_Q)-\vec{x}_Q(t'_Q))  d^3 x_{\Jnr}\nonumber \\
  & = & Q \left. \left\{ \left[\frac{\vec{\beta}_u\times 
   \hat{r}'}{(r')^2}+
       \frac{\partial ~}{\partial t}\left(\frac{\vec{\beta}_u\times 
   \hat{r}'}{c r'}\right)\right]\right\}\right|_{t' = t'_Q}.     
    \end{eqnarray}
    However, these formulae are not the ones obtained from (6.37) and (6.38) in Ref.~\cite{GH}.
    Instead, a change of variable is introduced into the $\delta$-functions in the first lines
    of (6.39) and (6.40):
   \begin{equation}
      \vec{z}(t'_Q) \equiv \vec{x}_{\Jnr}(t'_Q)-\vec{x}_Q(t'_Q).
   \end{equation}
       The Jacobian, $J$, relating the volume elements $d^3\vec{z}$ and
     $d^3\vec{x}_{\Jnr}$ according to 
   \begin{equation}
      d^3\vec{z} =  J d^3\vec{x}_{\Jnr}
      \end{equation}
      is introduced. It is then stated, without derivation, that
       $J = K$ where $K$ is Jacobian relating $dt$ to $dt'$, as given by Eqs.(5.15) and (6.3)
       above. The $x$-component of (6.41) is
  \begin{equation}
      z_x(t'_Q) = x_{\Jnr}(t'_Q)-x_Q(t'_Q).
   \end{equation}
     Since $x_Q(t'_Q)$ is constant it follows from (6.49) that $d z_x = dx_{\Jnr}$. Similarly.
     $d z_y = dy_{\Jnr}$  and $d z_z = dz_{\Jnr}$ so that the Jacobian $J$ in (6.42) is unity,
       not $K$, as claimed in Ref.~\cite{GH}.
       \par Since the authors of Ref.~\cite{GH} nevertheless {\it did} insert a factor $1/K$
        multiplying the $\delta$-functions in the first lines of (6.39) and (6.40), before performing
        the spatial integrations, the equations obtained were not those in the last lines 
        of (6.39) and (6.40) but instead:
      \begin{eqnarray}
  \vec{\Ei}({\rm GH2})^{ret}_{J = K} & = & \left.Q \left[\frac{\hat{r}'}{K(r')^2}+
       \frac{\partial ~}{\partial t}\left(\frac{\hat{r}'}{cKr'}\right)
       -  \frac{\partial ~}{\partial t}\left(\frac{[\vec{\beta}_u]}{cK r'}\right)\right]
     \right|_{t' = t'_Q},  \\
  \vec{\Bi}({\rm GH2})^{ret}_{J = K} & = & Q \left. \left\{ \left[\frac{\vec{\beta}_u \times 
   \hat{r}'}{K(r')^2}+
       \frac{\partial ~}{\partial t}\left(\frac{\vec{\beta}_u \times 
   \hat{r}'}{cK r'}\right)\right]\right\}\right|_{t' = t'_Q}     
    \end{eqnarray}
   where the subscript `$J = K$' distinguishes these equations from the formally correct
   ones (6.39) and (6.40), i.e. the correctly calculated point-like charge versions
    of the general equations (6.37) and (6.38), claimed to be the Jefimenko
    equations but actually given, without derivation, in Ref.~\cite{GH}. 
     \par After writing (6.44) and (6.45) (the equivalents, in gaussian units,
     of the MKS Eqs.(41) and (42) of  Ref~\cite{GH}) it is stated that they are
     `essentially in the form made famous by Feynman'. In fact Eq.~(6.44) is equivalent to
       Eq.~(19) of Ref.~\cite{JPS} on transforming the $t'$ derivatives in the latter into
    the $t$ derivatives of the former. It is correctly stated, but not demonstrated, in Ref.~\cite{JPS} that their  Eq.~(19) 
      is the same as the LW retarded electric field. 
      \par The introduction of the factor (1/K) inside the time derivatives in (6.44)
       and (6.45) is equivalent to replacing the retarded potentials in (6.22) and (6.23)
    by the LW potentials. It is shown in  Ref~\cite{GH} that (6.44) is equivalent to the LW
   electric field of Eq.~(6.6) and stated (without proof) that the magnetic field is given by the relation
 $\vec{\Bi}({\rm LW})^{ret} =   \hat{r}' \times  \vec{\Ei}({\rm LW})^{ret}$.
  \par Summarising the results obtained so far in this section: Ref.~\cite{JPS} does demonstrate the
    consistency of the Feynman 
    and LW formulae for retarded electric fields.
   The `derivation' of the Jefimenko formulae from the defining equations
    (5.1) and (5.2) of the electric and magnetic fields and the retarded potentials 
     (6.22) and (6.23) given in Ref.~\cite{GH} is erroneous due to mathematical misinterpretation
     of spatial partial derivatives. The same remark applies to
  Jefimenko's original derivation~\cite{Jefimenko} of these equations. 
   The Eqs.(6.44) and (6.45) given in Ref.~\cite{GH}
     are not the Jefimenko equations but are obtained from them by introducing an overall 
      multiplicative factor $1/K$ in each term and allowing the time derivatives to act
  on all factors in the terms of the equation instead of uniquely on the charge and
  current densities as in the Jefimenko formulae. This is tantamount to replacing the potentials 
  of (6.22) and (6.23) by the retarded LW potentials of Eq.~(2.28). Eq.(6.44) does give the LW field of (6.6),
    as claimed in Ref.~\cite{GH}.

   \par It was pointed out in Ref.~\cite{McDonald} that an equation for the magnetic
      field identical to the Jefimenko equation (6.14) and a formula equivalent to
      the Jefimenko electric field, (6.13), had been given earlier in the second
    edition of the book `Classical Electricity and Magnetism'
    by Panofsky and Phillips~\cite{PPJE}. The equivalent electric field formula is:
      \begin{equation}
      \vec{E}({\rm PP})^{ret} = \int \left\{\frac{\hat{r}'[\rho]}{(r')^2} +
       \frac{([\vec{\Ji}]\cdot \hat{r}')\hat{n}+([\vec{\Ji}]\times \hat{r}')\times \hat{r}'}{c(r')^2}
         +\frac{[\dot{\vec{\Ji}}]\times \hat{r}')\times \hat{r}'}{c^2 r'}\right\} d^3 x_{\Ji}.
      \end{equation}
      A calculation claiming to show the equivalence of (6.13) and (6.46) was given in Section II of  Ref.~\cite{McDonald}
     The first step was to repeat the erroneous derivation of (6.13) from the defining
      equation (5.1) of the electric field and the non-relativistic potentials (6.22) and (6.23),
      previously given in Ref.~\cite{GH} and discussed above. The term proportional
       to $\partial[\rho]/\partial t$ is manipulated to obtain Eq.(6.46). As shown above,
        this term actually vanishes.
 \SECTION{\bf{Summary}}
       Retarded potentials are derived from homogeneous d'Alembert
       equations for electromagnetic potentials and the Lorenz
       condition. The potentials so-obtained in Eq.(2.12) differ from the
       LW potentials of CEM. It is shown that the incorrect LW potentials result 
      from neglect of the dependence of the effective density
      of a moving charge distribution on its speed. This point is made
      particularly clear by the careful re-examination of Feynman's
      derivation of the LW potentials presented in Section 3. In
      Section 4, several `relativistic' derivations of the LW
      potentials or the corresponding retarded fields given in text
      books are reviewed. It is shown that they all contain
      misapplications of special relativity --in particular by
      invoking a spurious `length contraction' effect. In all
    of the relativistic derivations, retardation effects are 
    neglected, whereas in the original 19th Century derivations of the LW
    potentials or the corresponding retarded fields, no relativistic 
    effects are considered. There are therefore two independent,
    logically incompatible, and incorrect, derivations of retarded potentials
    and their associated fields. In Section 5 the retarded RCED fields of
    a uniformly moving charge are derived and expressed in `present
    time' form. Except for an overall multiplicative factor
    $1/(1-\hat{r}' \cdot \vec{\beta}_u)$ and the retarded time
    argument, they are the same as the instantaneous force fields
    of RCED~\cite{JHFRCED}. In Section 6 the
     consistency claimed in the pedagogical literature between various
     different formulae for the fields of an accelerated charge (LW,
     Feynman, Jefimenko) is considered. The Feynman formula for
    the retarded electric field of a charge in arbitary motion is
    (as previously shown in Ref.~\cite{JPS}) consistent with the LW
     field. The electric field of a
    uniformly moving charge given by the Jefimenko formula is found to
    be, unlike CEM and RCED fields, Coulombic. 
   \par The considerations of the present paper are of a primarily
    mathematical nature. The physical interpretation of retarded
    (radiation) and instantaneous (force) fields in RCED has been
    discussed in some detail previously~\cite{JHFRCED,JHFFT}.
 
\newpage
 \par{\bf Appendix A}
\renewcommand{\theequation}{A.\arabic{equation}}
\setcounter{equation}{0}
 \par In this appendix, retarded electric and magnetic fields are derived from the LW potentials
  as well as from the equations (6.44) and (6.45) equivalent to those given in Ref~\cite{GH} and claimed there
  to be the same as the LW fields. To derive the LW fields the potentials
  \begin{eqnarray}
    \Ai_0({\rm LW})^{ret} & = & \left.\frac{Q}{K r'}\right|_{t' = t'_Q}, \\
 \vec{\Ai}({\rm LW})^{ret} & = & \left.\frac{\vec{\beta}_u}{K r'}\right|_{t' = t'_Q}
  \end{eqnarray}
  where $K \equiv (1-\hat{r}' \cdot \vec{\beta}_u)$, are substituted into the defining equations
   (5.1) and (5.2) of electric and magnetic fields to give
  \begin{eqnarray}
    \vec{\Ei}({\rm LW})^{ret}& = & -\vec{\nabla} \Ai_0({\rm LW})^{ret} -
      \frac{1}{c} \frac{\partial \vec{\Ai}({\rm LW})^{ret}}{\partial t}, \\
 \vec{\Bi}({\rm LW})^{ret} & = & \vec{\nabla} \times  \vec{\Ai}({\rm LW})^{ret}.
  \end{eqnarray}
  For simplicity, all labels, superscripts and subscripts on the fields and potentials are omitted in the
   following.
  \par Taking into account, by the chain rule, the contribution to the fields of each 
        factor in the potentials, (A.3) and (A.4) give:
   \begin{eqnarray}  
   \vec{\Ei}  & = & -\frac{Q}{K}\vec{\nabla}\left(\frac{1}{r'}\right)-
     \frac{Q}{r'}\vec{\nabla}\left(\frac{1}{K}\right)  
   -\frac{Q \vec{\beta}_u}{c K} \frac{\partial ~}{\partial t}\left(\frac{1}{r'}\right)
    -\frac{Q \vec{\beta}_u}{c r'} \frac{\partial ~}{\partial t}\left(\frac{1}{K}\right)
    -\frac{Q}{c K r'} \frac{\partial \vec{\beta}_u }{\partial t}, \\
   \vec{\Bi}  & = & -\frac{Q}{K}\vec{\beta}_u \times\vec{\nabla}\left(\frac{1}{r'}\right)
 -\frac{Q}{r'}\vec{\beta}_u \times\vec{\nabla}\left(\frac{1}{K}\right)
   + \frac{Q}{K r'}(\vec{\nabla} \times \vec{\beta}_u). 
 \end{eqnarray}
   In these and the following equations it is understood that all spatial partial
    derivatives hold $t$ constant and all temporal partial deivatives hold $\vec{x}_q$,
    the field position, constant. The derivatives in the successive terms on the right sides 
   of these equations are now evaluated. 
   \par The first term on the right side of (A.5) gives: 
   \begin{eqnarray}  
   -\frac{Q}{K}\vec{\nabla}\left(\frac{1}{r'}\right) & = & -\frac{ \hat{\imath} Q}{K}
    \frac{\partial~}{\partial x_q}\left(\frac{1}{r'}\right)+ .~.~.~. \nonumber \\
     & = &  \frac{ \hat{\imath} Q}{K(r')^2}
    \frac{\partial r'}{\partial x_q}+ .~.~.~. \nonumber \\
     & = &  \frac{ \hat{\imath} Q (x_q-x_Q)}{K^2(r')^3}+ .~.~.~. \nonumber \\
     & = &  \frac{ \hat{r'}}{K^2(r')^2}
 \end{eqnarray}
    where Eq.(5.8) has been used.
   \par Considering the second term on the right side of (A.5),
  \begin{eqnarray} 
  -\frac{Q}{r'}\vec{\nabla}\left(\frac{1}{K}\right) & = &   -\frac{ \hat{\imath} Q}{ r' K^2}
    \frac{\partial~}{\partial x_q}(\hat{r}' \cdot \vec{\beta}_u)+ .~.~.~ \nonumber \\
 & = &   -\frac{ \hat{\imath} Q}{ r' K^2}
   \left[\vec{\beta}_u \cdot \frac{\partial \hat{r}'}{\partial x_q}
     + \hat{r}' \cdot \frac{\partial \vec{\beta}_u }{\partial x_q}\right]+ .~.~.~ \nonumber \\
 & = &   -\frac{ \hat{\imath} Q}{ r' K^2}
   \left[\vec{\beta}_u \cdot \frac{\partial \hat{r}'}{\partial x_q}
   + \frac{\partial t'}{\partial x_q}(\hat{r}'\dot{\cdot \vec{\beta}_u})\right]+ .~.~.~~.
 \end{eqnarray}
  Now,
   \begin{eqnarray} 
    \frac{\partial \hat{r}'}{\partial x_q} & = &  \frac{\partial ~}{\partial x_q}        
    \left(\frac{\vec{r}'}{r'}\right) = \frac{1}{r'} \frac{\partial \vec{r}'}{\partial x_q}
     - \frac{\vec{r}'}{(r')^2} \frac{\partial r'}{\partial x_q}\nonumber \\
   & = & \frac{\hat{\imath}}{r'}\left(1- \frac{d x_Q}{d t'}\frac{\partial t'}{\partial x_q}\right)
      - \frac{\hat{r}'}{r'} \frac{\partial r'}{\partial x_q}\nonumber \\
    & = & \frac{\hat{\imath}}{r'} + \frac{\hat{\imath} \beta_u -\hat{r}'}{r'} 
    \frac{\partial r'}{\partial x_q}\nonumber \\
  & = & \frac{\hat{\imath}}{r'} + \frac{(\hat{\imath} \beta_u -\hat{r}')(x_q-x_Q)}{K(r')^2} 
  \end{eqnarray}
  where Eqs.(5.7) and (5.8) have been used, as well as the assumption that $\vec{u}$ is
   parallel to the $x$-axis. Substituting (A.9) in (A.8) and again using Eqs.(5.7) and (5.8)
   gives:
   \begin{eqnarray}
   -\frac{Q}{r'}\vec{\nabla}\left(\frac{1}{K}\right) & = & -\frac{Q \vec{\beta}_u}{K^2(r')^2}
       -Q \hat{\imath}\frac{[\beta_u^2-(\hat{r}' \cdot \vec{\beta}_u)]}{K^3(r')^2}
        \frac{(x_q-x_Q)}{r'}+ Q \hat{\imath}\frac{(\hat{r}' \cdot \dot{\vec{\beta}_u})]}{c K^3r'}
        \frac{(x_q-x_Q)}{r'}+.~.~.~. \nonumber \\
         & = & -\frac{Q \vec{\beta}_u}{K^2(r')^2}
       -Q\hat{r}'\frac{[\beta_u^2-(\hat{r}' \cdot \vec{\beta}_u)]}{K^3(r')^2}
        + Q \hat{r}'\frac{(\hat{r}' \cdot \dot{\vec{\beta}_u})]}{c K^3r'}+.~.~.~.~~.
  \end{eqnarray}                       
   The third term on the right side of (A.5) gives
  \begin{eqnarray}
   -\frac{Q \vec{\beta}_u}{c K} \frac{\partial ~}{\partial t}\left(\frac{1}{r'}\right) & = &
     -\frac{Q\vec{\beta}_u}{c K} \frac{\partial t'}
  {\partial t} \frac{\partial ~}{\partial t'}\left(\frac{1}{r'}\right) \nonumber \\
  & = & \frac{Q \vec{\beta}_u}{c K^2(r')^2} \frac{\partial r'}
  {\partial t'} \nonumber \\
  & = & -\frac{Q \vec{\beta}_u (\hat{r}' \cdot \vec{\beta}_u)}{K^2(r')^2}
  \end{eqnarray}
   where (5.15) and (5.13) have been used.
  The fourth term on the right side of (A.5) gives
   \begin{equation}
  -\frac{Q \vec{\beta}_u}{c r'} \frac{\partial ~}{\partial t}\left(\frac{1}{K}\right)
     = -\frac{Q \vec{\beta}_u}{c r'} \frac{\partial t'} {\partial t} 
     \frac{\partial ~}{\partial t'}\left(\frac{1}{K}\right)
    \end{equation}
   But
   \begin{equation}
  \frac{\partial ~}{\partial t'}\left(\frac{1}{K}\right) = \frac{\partial ~}{\partial t'}
    \left(\frac{1}{1- (\hat{r}' \cdot \vec{\beta}_u)}\right) =
    \frac{1}{K^2}\frac{\partial(\hat{r}' \cdot \vec{\beta}_u)}{\partial t'} =
     \frac{1}{K^2}\left[\vec{\beta}_u \cdot \frac{\partial \hat{r}'}{\partial t'}
     +\hat{r}' \cdot \dot{\vec{\beta}_u}\right]
      \end{equation}
  and also
 \begin{equation}
  \frac{\partial \hat{r}'}{\partial t'} = \frac{\partial ~}{\partial t'}\left(\frac{\vec{r}'}{r'}\right)
     = -c\frac{\vec{\beta}_u}{r'}-\frac{\vec{r}'}{(r')^2}\frac{\partial r'}{\partial t'}
    = c\frac{[ \hat{r}'(\hat{r}' \cdot \vec{\beta}_u)-\vec{\beta}_u]}{r'}
   \end{equation}
  where (5.13) has been used.
   (A.12)-(A.14) together with (5.15) then give
 \begin{equation}
   -\frac{Q \vec{\beta}_u}{c r'} \frac{\partial ~}{\partial t}\left(\frac{1}{K}\right)
     =  -\frac{Q \vec{\beta}_u}{K^3}\left\{ \frac{(\hat{r}' \cdot \vec{\beta}_u)^2-\beta_u^2)}{(r')^2}
      + \frac{\hat{r}' \cdot \dot{\vec{\beta}_u}}{c r'}\right\}.
  \end{equation}
  Collecting together (A.7), (A.10), (A.11) and (A.15) gives, for the electric field derived 
    from the LW potentials:
    \begin{eqnarray} 
     \vec{\Ei} & = & \frac{Q}{K^2(r')^2}\left\{\hat{r}'- \vec{\beta}_u[1+ (\hat{r}' \cdot \vec{\beta}_u)]
              -\frac{\vec{\beta}_u[(\hat{r}' \cdot \vec{\beta}_u)^2-\beta_u^2]+\hat{r}'
      [ \hat{r}' \cdot \vec{\beta}_u-\beta_u^2]}{K}\right\}  \nonumber \\
  &   & - \frac{Q}{K^2 c r'}\left[\dot{\vec{\beta}_u}
        +\frac{(\vec{\beta}_u -\hat{r}')\hat{r}' \cdot \dot{\vec{\beta}_u}}{K}\right] \nonumber \\
      & = & \frac{Q}{K^3}\left[\frac{\hat{r}'-\vec{\beta}_u}{\gamma_u^2(r')^2}
       + \frac{{\hat{r}' \times [(\hat{r}'-\beta}_u) \times\dot{\vec{\beta}_u}]}{c r'} \right]. 
    \end{eqnarray}
 
  \par  The retarded LW magnetic field given by (A.6) is now calculated.
   \par Since $\vec{u}$ is assumed to be parallel to the $x$-axis, it follows that
       \begin{eqnarray} 
 -\frac{Q}{K}\vec{\beta}_u \times\vec{\nabla}\left(\frac{1}{r'}\right) & = &
  -\frac{Q\hat{k} \beta_u}{K}\frac{\partial ~}{\partial y_q}\left(\frac{1}{r'}\right) \nonumber \\ 
     & = & \frac{Q\hat{k} \beta_u}{K(r')^2}\frac{\partial r'}{\partial y_q}
    = \frac{Q\hat{k} \beta_u y_q}{K^2(r')^3} \nonumber \\
    & = &   \frac{Q(\vec{\beta}_u \times \hat{r}')}{K^2(r')^2}. 
 \end{eqnarray}
  Similarly
       \begin{eqnarray} 
 -\frac{Q}{K}\vec{\beta}_u \times\vec{\nabla}\left(\frac{1}{K}\right) & = &
  -\frac{Q\hat{k} \beta_u}{K}\frac{\partial ~}{\partial y_q}\left(\frac{1}{K}\right) \nonumber \\ 
     & = & -\frac{Q\hat{k} \beta_u}{K^2 r'}\frac{\partial (\hat{r}' \cdot \vec{\beta}_u)}{\partial y_q}
 \nonumber \\
      & = & -\frac{Q \hat{k} \beta_u}{K^2r'} \left(\hat{r}' \cdot \frac{\partial \vec{\beta}_u}{\partial y_q}+\vec{\beta}_u 
  \cdot \frac{\partial \hat{r}'}{\partial y_q} \right).
 \end{eqnarray}
    Evaluating the first term in brackets on the right side of Eq.(A.18),
\begin{equation}
   \hat{r}' \cdot \frac{\partial \vec{\beta}_u}{\partial y_q}
   =  \frac{\partial t'}{\partial y_q}(\hat{r}' \cdot \dot{\vec{\beta}_u})
       = -\frac{y_q(\hat{r}' \cdot \dot{\vec{\beta}_u})}{cK r'}
 \end{equation}
      where the relation 
\begin{equation}
 \frac{\partial t'}{\partial y_q} = -\frac{1}{c}  \frac{\partial r'}{\partial y_q}
 \end{equation}  
 given by differentiating the retardation condition $t' = t-r'/c$ as well as Eq.(5.18) 
 have been used.
  The second tern in brackets on the right side of(A.18) is 
\begin{equation}
   \vec{\beta}_u  \cdot \frac{\partial \hat{r}'}{\partial y_q}
    = \vec{\beta}_u  \cdot \frac{\partial ~}{\partial y_q} \left(\frac{\vec{r}'}{r'}\right)
    =  \vec{\beta}_u  \cdot \left( \frac{1}{r'} \frac{\partial \vec{r}'}{\partial y_q}-
       \frac{\vec{r}'}{(r')^2 }\frac{\partial r'}{\partial y_q} \right).
 \end{equation}
     Assuming, without  loss of generality, that the vector $\vec{r}'$ is confined to the
     $x-y$ plane,
\begin{equation}
  \frac{\partial \vec{r}'}{\partial y_q} = -\hat{\imath}\frac{d x_Q}{d t'}\frac{\partial t'}{\partial y_q}
                  + \hat{\jmath}.
 \end{equation}
    Combining (A.21) and (A.22), again using (A.20) and (5.18), gives
\begin{equation}
  \vec{\beta}_u  \cdot \frac{\partial \hat{r}'}{\partial y_q} = \frac{[\beta_u^2-
   \hat{r}' \cdot \vec{\beta}_u]y_q}{K(r')^2}.
 \end{equation}
   Combining A(18), (A.19) and (A.23),
\begin{equation}
   -\frac{Q}{K}\vec{\beta}_u \times\vec{\nabla}\left(\frac{1}{K}\right) = 
     \frac{Q \vec{\beta}_u \times \hat{r}'}{K^3}\left[ \frac{[ \hat{r}' \cdot \vec{\beta}_u- \beta_u^2]}{(r')^2}
       +\frac{\hat{r}' \cdot \dot{\vec{\beta}_u}}{cr'} \right]
 \end{equation}
     The third term on the right side of (A.6) is
  \begin{eqnarray}
   \frac{Q}{K r'}(\vec{\nabla} \times \vec{\beta}_u) & = &   -\frac{Q \hat{k}}{K r'} \frac{\partial \beta_u}{\partial y_q}
     =  -\frac{Q \hat{k}}{K r'} \frac{\partial t'}{\partial y_q} \dot{\beta}_u \nonumber \\
    & = &  \frac{Q \hat{k}}{c K r'} \frac{\partial r'}{\partial y_q} \dot{\beta}_u 
         =  \frac{Q (\vec{\beta}_u \times \hat{r}')}{c K^2 r'\beta_u}\dot{\beta}_u
    \end{eqnarray}
   where (A.20) and (5.18) have been used. 
  \par Collecting together (A.17), (A.24) and  (A.25), the magnetic field generated by the LW 
   potentials is:
     \begin{eqnarray}
   \vec{\Bi} & = & \left. \left\{\frac{Q (\vec{\beta}_u \times \hat{r}')}{K^3}\left[\frac{K + \hat{r}' \cdot \vec{\beta}_u
 -\beta_u^2}{(r')^2} +\frac{K \dot{\beta}_u+ \beta_u(\hat{r}' \cdot \dot{\vec{\beta}_u})}{c r' \beta_u}\right] \right\}
     \right|_{t' = t'_Q},  \nonumber \\
 & = & \left. \left\{\frac{Q (\vec{\beta}_u \times \hat{r}')}{K^3}\left[\frac{1}{\gamma_u^2(r')^2}
   +\frac{(\dot{\beta}_u(1-\hat{r}' \cdot \vec{\beta}_u)
 + \beta_u(\hat{r}' \cdot \dot{\vec{\beta}_u})}{c r' \beta_u}\right]
   \right\} \right|_{t' = t'_Q}.
   \end{eqnarray}
 (A.16) and (A.26) are the formulae (6.6) and (6.7) of the text.
  \par The consistency of the fields of Eqs.(6.44) and (6.45) with the LW fields
        of (6.6) and (6.7) claimed in Ref.~\cite{GH} is now investigated. The equations
    analogous to (A.5) and (A.6) given by using the chain rule to expand the derivatives in
  (6.44) and (6.45) are:
  \begin{eqnarray}  
   \vec{\Ei}  & = & \frac{Q \hat{r}'}{K (r')^2} +  \frac{Q \hat{r}'}{c r'}
   \frac{\partial ~}{\partial t}\left(\frac{1}{K}\right) + 
   \frac{Q \hat{r}'}{c K} \frac{\partial ~}{\partial t}\left(\frac{1}{r'}\right)+
 \frac{Q}{c K r'}\frac{\partial  \hat{r}'}{\partial t}
    \nonumber \\  
  &  &  -\frac{Q \vec{\beta}_u}{c K} \frac{\partial ~}{\partial t}\left(\frac{1}{r'}\right)
    -\frac{Q \vec{\beta}_u}{c r'} \frac{\partial ~}{\partial t}\left(\frac{1}{K}\right)
    -\frac{Q}{c K r'} \frac{\partial \vec{\beta}_u }{\partial t}, \\
   \vec{\Bi}  & = &  \frac{Q (\vec{\beta}_u \times \hat{r}')}{K (r')^2}+ 
    \frac{Q(\vec{\beta}_u \times \hat{r}')}{c r'}
   \frac{\partial ~}{\partial t}\left(\frac{1}{K}\right)+
  \frac{Q (\vec{\beta}_u \times \hat{r}')}{c K}
  \frac{\partial ~}{\partial t}\left(\frac{1}{r'}\right)   \nonumber \\ 
 &  & +\frac{Q}{c K r'}\left(\vec{\beta}_u \times\frac{\partial  \hat{r}'}{\partial t}\right)-\frac{Q}{c K r'}
 \left( \hat{r}' \times \frac{\partial \vec{\beta}_u }{\partial t}\right).
 \end{eqnarray}
 .  Comparison with (A.5) and (A.6) shows that the last three terms in (A.5) and (A.27) (originating
     from the time derivative in (A.3)) are the same but all other terms in (A.27) and (A.28)
     differ from those in (A.5) and (A.6). Thus to compare (A.5) and (A.27) the derivatives in the
     second. third and fourth terms on the right of (A.27)  must be evaluated, while to compare 
     (A.6) and (A.28) all derivatives on the right side of (A.28) must be evaluated. This is readily
      done using, {\it mutatis mutandis}, the formulae obtained above.
    \par The second term in (A.27) is
 \begin{equation}
  \frac{Q \hat{r}'}{c r'}
   \frac{\partial ~}{\partial t}\left(\frac{1}{K}\right) =
   \frac{Q \hat{r}'}{c K r'}
   \frac{\partial ~}{\partial t'}\left(\frac{1}{K}\right) =
     \frac{Q \hat{r}'}{K^3}\left\{ \frac{(\hat{r}' \cdot \vec{\beta}_u)^2-\beta_u^2)}{(r')^2}
      + \frac{\hat{r}' \cdot \dot{\vec{\beta}_u}}{c r'}\right\}
   \end{equation}
  by analogy with Eq.(A.15). 
    \par The third term is
  \begin{equation}   
  \frac{Q \hat{r}'}{c K}\frac{\partial ~}{\partial t}\left(\frac{1}{r'}\right)
      =   \frac{Q \hat{r}'}{c K^2}\frac{\partial ~}{\partial t'}\left(\frac{1}{r'}\right)
      = -\frac{Q \hat{r}'}{c K^2 (r')^2} \frac{\partial r'}{\partial t'}
      =  \frac{Q \hat{r}' (\hat{r}' \cdot \vec{\beta}_u)}{K^2 (r')^2}
 \end{equation}
    where (5.15) and (5.13) have been used
  \par The fourth term is
 \begin{equation} 
 \frac{Q}{c K r'}\frac{\partial  \hat{r}'}{\partial t}
   = \frac{Q}{c K^2 r'}\frac{\partial  \hat{r}'}{\partial t'}
   =\frac{Q [ \hat{r}'(\hat{r}' \cdot \vec{\beta}_u)-\vec{\beta}_u]}{K^2(r')^2}.
 \end{equation}
   Substituting (A.29)-(A.31) into (A.22) as well as the previously obtained terms, and performing
   some algebraic simplification, gives
 \begin{equation} 
   \vec{\Ei}  =  \frac{Q}{K^3}\left[\frac{\hat{r}'-\vec{\beta}_u}{\gamma_u^2(r')^2}
       + \frac{{\hat{r}' \times [(\hat{r}'-\beta}_u) \times\dot{\vec{\beta}_u}]}{c r'} \right]. 
   \end{equation}  
     Which  is the LW field of Eq.(A.16)
 
  Noting that the first three terms of (A.28) differ from those of (A.27) by the replacement
  $\hat{r}' \rightarrow \vec{\beta}_u \times \hat{r}'$ and using the above results for the
   latter terms, gives, after algebraic simplification, the LW magnetic field of Eq.~(6.7). 
 \newpage
 \par{\bf Appendix B}
\renewcommand{\theequation}{B.\arabic{equation}}
\setcounter{equation}{0}
  \par For clarity the total time derivatives in (6.6) are replaced the corresponding partial derivatives
   with respect to the present time, $t$, for a fixed value of the field point position $\vec{x}_q$.
   \par The second term on the right side of (6.10) contains the derivative:
  \begin{equation}
  \frac{\partial~}{\partial t}\left(\frac{\hat{r}'}{(r')^2}\right)
     = \frac{\partial t'}{\partial t} \frac{\partial~}{\partial t'}
  \left(\frac{\vec{r}'}{(r')^3}\right) =
    \frac{\partial t'}{\partial t}\left[\frac{1}{(r')^3} \frac{\partial \vec{r}'}{\partial t'}
 -\frac{3  \vec{r}'}{(r')^4}\frac{\partial r'}{\partial t'}\right].
  \end{equation}
   Since the vector $\vec{r}'$ is confined to the $x$-$y$ plane,
   \begin{equation}
 \frac{\partial \vec{r}'}{\partial t'} = \hat{\imath} \frac{\partial (x_q-x_Q)}{\partial t'}
       + \hat{\jmath} \frac{\partial y_q}{\partial t'} = -c\vec{\beta}_u
 \end{equation}
    since $\partial y_q/\partial t' = 0$ and $c \beta_u = d x_Q/dt$.
     (5.13), (5.15), (B.1) and (B.2) give: 
\begin{equation}
  \frac{r'}{c}\frac{\partial~}{\partial t}\left(\frac{\hat{r}'}{(r')^2}\right)
  = \frac{1}{1-\hat{r}' \cdot \vec{\beta}_u}\left[ \frac{3 \hat{r}'(  \hat{r}' \cdot \vec{\beta}_u)
        - \vec{\beta}_u}{(r')^2}\right].
 \end{equation}
 \par Considering now the last term on the right side of (6.10):
\begin{equation}
  \frac{\partial^2 \hat{r}'}{\partial t^2} = \frac{\partial t'}{\partial t}\frac{\partial~}{\partial t'}
    \left[\frac{\partial t'}{\partial t}\frac{\partial \hat{r}'}{\partial t'}\right]
    =  \frac{\partial t'}{\partial t}\left[\frac{\partial \hat{r}'}{\partial t'} \frac{\partial~}{\partial t'}
      \left(\frac{\partial t'}{\partial t}\right)+ \frac{\partial t'}{\partial t}
      \frac{\partial^2  \hat{r}'}{\partial t'^2}\right]
 \end{equation}
   where
     \begin{eqnarray}
     \frac{\partial~}{\partial t'}\left(\frac{\partial t'}{\partial t}\right) & = &
    \frac{\partial~}{\partial t'}\left(\frac{1}{1-\hat{r}' \cdot \vec{\beta}_u}\right) =
     \frac{1}{(1-\hat{r}' \cdot \vec{\beta}_u)^2} \frac{\partial(\hat{r}' \cdot \vec{\beta}_u)}{\partial t'}
       \nonumber \\
    & = &  \frac{1}{(1-\hat{r}' \cdot \vec{\beta}_u)^2}\left[ c \frac{[(\hat{r}' \cdot \vec{\beta}_u)^2 -\beta_u^2]}
        {r'}+ (\hat{r}' \cdot \dot{\vec{\beta}_u)}\right]
      \end{eqnarray} 
    where (A.14) has been used.
      It also follows from (A.14) that
     \begin{eqnarray}
   \frac{\partial^2  \hat{r}'}{\partial t'^2} & = &
     - \frac{c[\hat{r}'(\hat{r}' \cdot \vec{\beta}_u)-\vec{\beta}_u]}{(r')^2}\frac{\partial r'}{\partial t'}
        +\frac{c}{r'}\left[\left(\frac{\partial \hat{r}'}{\partial t'}\right)(\hat{r}' \cdot \vec{\beta}_u)
        + \hat{r}'\left(\frac{\partial \hat{r}'}{\partial t'}\right) \cdot  \vec{\beta}_u
       + \hat{r}'(\hat{r}' \cdot \dot{\vec{\beta}_u})- \dot{\vec{\beta}_u}\right] \nonumber \\
      & = & c^2\left[\frac{\hat{r}'[3(\hat{r}' \cdot \vec{\beta}_u)^2-\beta_u^2]
      - 2\vec{\beta}_u(\hat{r}' \cdot \vec{\beta}_u)}{(r')^2}\right]
       + \frac{ c[\hat{r}'(\hat{r}' \cdot \dot{\vec{\beta}_u})-\dot{\vec{\beta}_u}]}{r'}.
     \end{eqnarray}
      Combining (5.15) and (A.14),(B.5) and (B.6) then gives
       \begin{eqnarray}
        \frac{1}{c^2} \frac{\partial^2  \hat{r}'}{\partial t^2} & = & \frac{1}{(1-\hat{r}' \cdot \vec{\beta}_u)}
   \left\{\frac{[\hat{r}'(\hat{r}'\cdot\vec{\beta}_u)- \vec{\beta}_u]}{(1-\hat{r}' \cdot \vec{\beta}_u)^2}
   \left[\frac{\hat{r}'(\hat{r}'\cdot \vec{\beta}_u)^2-\beta_u^2}{(r')^2}+\frac{\hat{r}' \cdot \dot{\vec{\beta}_u}}
        {c r'}\right]\right. \nonumber  \\
    &+&\left. \frac{1}{(1-\hat{r}' \cdot \vec{\beta}_u)}\left[\frac{\hat{r}'[3(\hat{r}' \cdot \vec{\beta}_u)^2
      -\beta_u^2] -2\vec{\beta}_u(\hat{r}'\cdot \vec{\beta}_u)}{(r')^2}+
       \frac{\hat{r}'( \hat{r}' \cdot \dot{\vec{\beta}_u})- \dot{\vec{\beta}_u}}{c r'}\right]\right\}.
      \end{eqnarray}
      Inserting (B.3) and (B.7) into (6.10) yields Eq.(6.12) of the text.
 
\end{document}